\documentclass[10pt, journal, a4paper, final, twoside, nofonttune]{IEEEtran}
\usepackage{amssymb,amsmath,amsthm,epsfig,graphics,float,subfigure,cite,mhchem}
\usepackage[colorlinks=true, allcolors=blue]{hyperref}
\usepackage{nicematrix}

\usepackage[english]{babel}

\begin{document}

\title{Biophysical aspects of neurocognitive modeling\\ with long-term sustained temperature variations}
\author{S.~Kernbach, O.~Kernbach, A.~Kernbach\\
\vspace{2mm}
\small CYBRES GmbH, Research Center of Advanced Robotics and Environmental Science,\\ Melunerstr. 40, 70569 Stuttgart, Germany, contact author: {\textit{serge.kernbach@cybertronica.de.com}}
\vspace{-6mm}}
\date{}
\maketitle
\thispagestyle{empty}

\begin{abstract}
Long-term focused attention with visualization and breathing exercises is at the core of various Eastern traditions. Neurocognitive and psychosomatic phenomena demonstrated during such exercises were instrumentally explored with EEG and other sensors. Neurocognitive modeling in the form of meditative visualization produced persistent temperature effects in the body long after the exercise finished; this raises the question about their psychosomatic or biophysical origin. The work explores this question by comparing experiments with focusing attention inside and outside the body. EEG, temperature, heart and breathing sensors monitor internal body conditions, high resolution differential calorimetric sensors are used to detect thermal effects outside the body. Experiments with 159 attempts (2427 operator-sensor sessions) were carried over five months, control measurements run in the same conditions in parallel to experimental series. Increase of body temperature up to moderate fever zone 38.5 $^\circ$C and intentional control of up and down trend of core temperature by 1.6 $^\circ$C are demonstrated. Persistent temperature variations last $>$60 min. Experiments also demonstrated induced thermal fluctuations at $10^{-3}$ $^\circ$C level in external calorimetric systems with 15 ml of water for 60-90 min. Repeatability of these attempts is over 90\%, statistical Chi-square and Mann-Whitney tests reject the null hypotheses about random character of outcomes. Thus, the obtained data confirm the persistent thermal effects reported in previous publications and indicate their biophysical dimension. To explain these results we refer to a new model in neuroscience that involves spin phenomena in biochemical and physical systems. These experiments demonstrate complex biophysical mechanisms of altered states of consciousness; their function in the body's neurohumoral regulation and non-classical brain functions is discussed.
\end{abstract}


\section{Introduction}

Neurocognitive and psychosomatic mechanisms of altered states of consciousness (ASC) in several Eastern traditions, such as Reiki \cite{baldwin2008reiki}, Tai chi \cite{chu2004tai}, Qigong \cite{pierce2007use}, Johrei \cite{gasiorowska2009clinical} and others have been studied in academic publications. For instance, the focused attention on the body (abdominal region near the navel) in yoga Tummo has been shown to increase core temperature up to 38.3$^\circ$C \cite{Kozhevnikov13}. Authors demonstrated that a neurocognitive modeling in the form of meditative visualization in ASC is responsible for two effects: (a) increasing temperature far beyond the range of normal body temperature; (b) persistent, step-wise temperature increase long after the exercises finished. Similar effects were described in other publications \cite{Gurtovoi92en,Godik10}.

Persistent effects were also reported for techniques targeting environmental objects and processes, such as exposure of external Qi \cite{Xin11,Chen04,Fukushima01} and focused attention \cite{Korenbaum21, Matos17} to aqueous solutions and properties of materials \cite{Xin99}, bacterial cultures and laboratory animals \cite{rubik2006vitro}. The practice of these techniques causes instrumentally recorded persistent changes in both the internal metabolism of human meditator and the properties of materials and processes. Comparing both results, the question arises about only psychosomatic or also biophysical origin of such phenomena. The work explorers this question in relation to long-term sustained temperature variations in two cases, when the attention is focused inside and outside the body. 

This experimental study replicates and extends the methodologies of previous works \cite{Kozhevnikov13, Gurtovoi92en, Xin99} with practitioners of the six yogas of Naropa \cite{Tsongkhapa14,Mullin97} and the Dhammakaya meditation \cite{SUDSUANG1991543}. Skin surface and core temperature are measured with IR camera (wavelength 8-14$\mu$m) and contact sensors. External effects are measured by high resolution differential calorimeters with 6 temperature sensors and 2x15 ml of water. Such calorimeters are able to measure extremely small \textit{t} changes as a symmetry breaking effect. A separate laboratory was prepared for measurements, three calorimetric setups served as control systems and three others used in experiments, all systems running in parallel. Additionally, the power supply, electromagnetic emissions, magnetic fields, air pressure and mechanical impacts are recorded by corresponding sensors. Meditators use the real-time biofeedback with sensors of EEG, heart and respiratory rate, and muscle activity.

The obtained results confirm the persistent effects reported in \cite{Kozhevnikov13, Gurtovoi92en, Xin99}. In particular, increase of body temperature up to moderate fever zone 38.5 $^\circ$C and intentional control of up and down trend of core temperature by 1.6 $^\circ$C are demonstrated for $>$60 min. Experiments have also shown the induced thermal fluctuations in external calorimetric systems at $10^{-3}$ $^\circ$C level for 60-90 min, which point to biophysical dimension of such phenomena. To explain these results, the biophysical hypothesis consists in the spin conversion of water isomers \cite{Pershin15Biophysics}. The spin configuration affects chemical reactivity \cite{Kilaj18,kernbach2022electrochemical}, behavior in electric field \cite{Horke:192659}, magnetic moment \cite{doi:10.1126/science.1200433}, evaporation \cite{POULOSE2023814}, surface tension \cite{kernbach2023Pershin} and other properties, in particular, the heat capacity of water isomers \cite{kernbach23Thermal}. As a hypothesis, it is assumed that such a spin conversion induced through ASC, as recently proposed in neuroscience \cite{HU2004633, Chen20} e.g. via non-classical brain functions \cite{Kerskens_2022}, can cause the observed thermal effects.

\section{Setup and methodology of experiments}

\subsection{Description of the setup}

The experimental setup includes biofeedback equipment for monitoring body-internal conditions and differential calorimeters for experiments with external thermal phenomena. 
\begin{figure}[ht]
\centering
\subfigure[]{\includegraphics[width=0.45\textwidth]{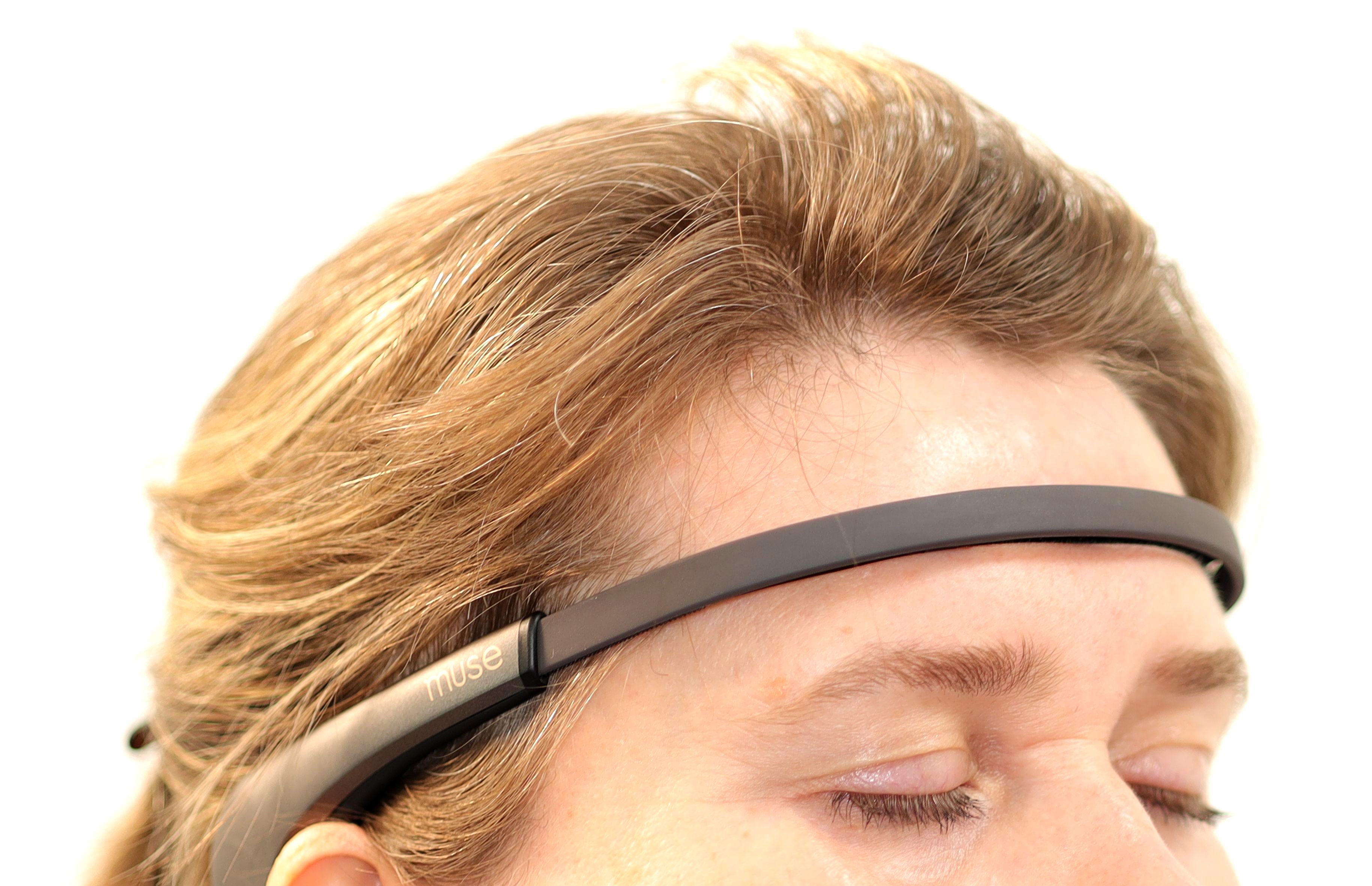}}
\subfigure[]{\includegraphics[width=0.45\textwidth]{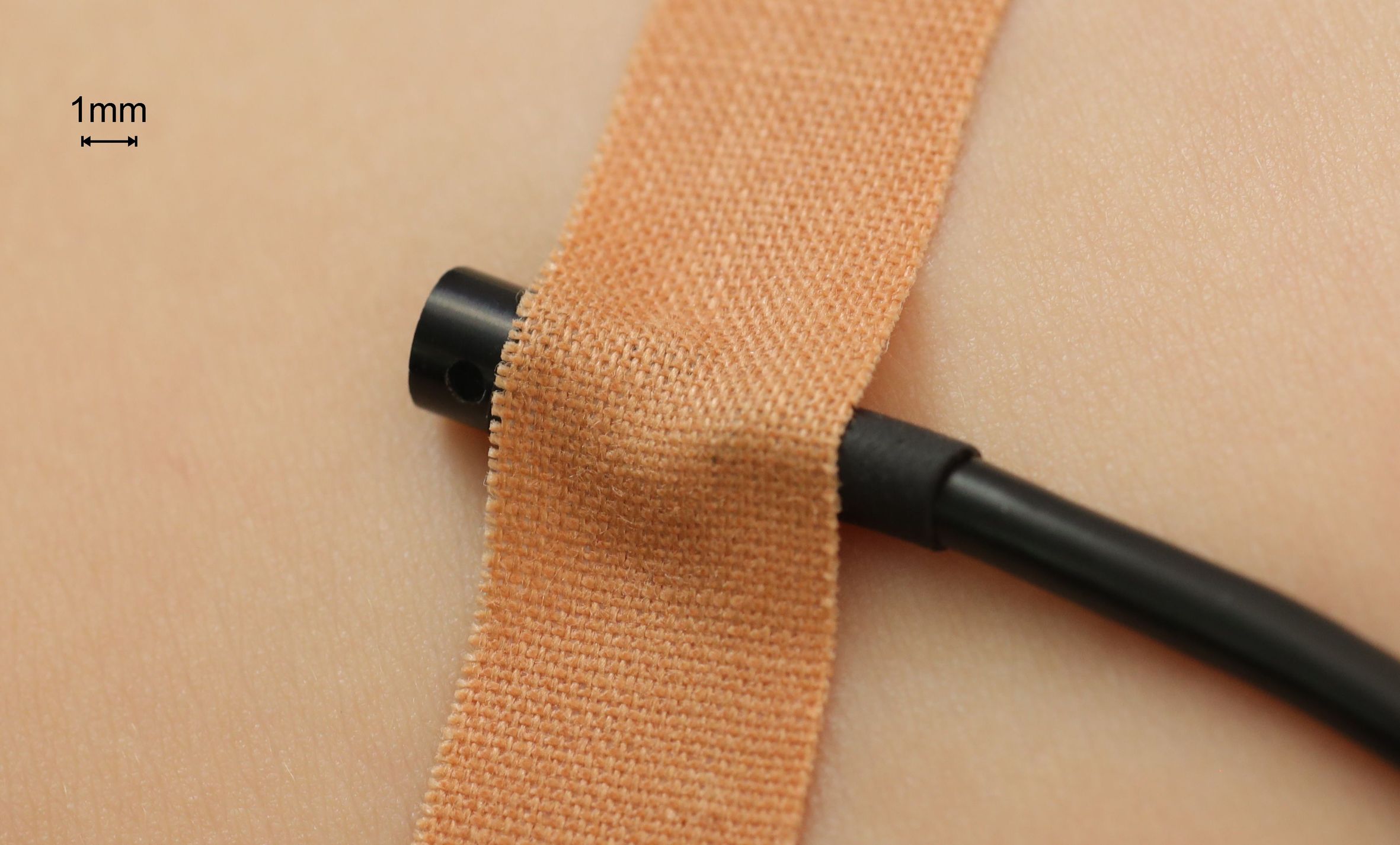}}
\subfigure[]{\includegraphics[width=0.45\textwidth]{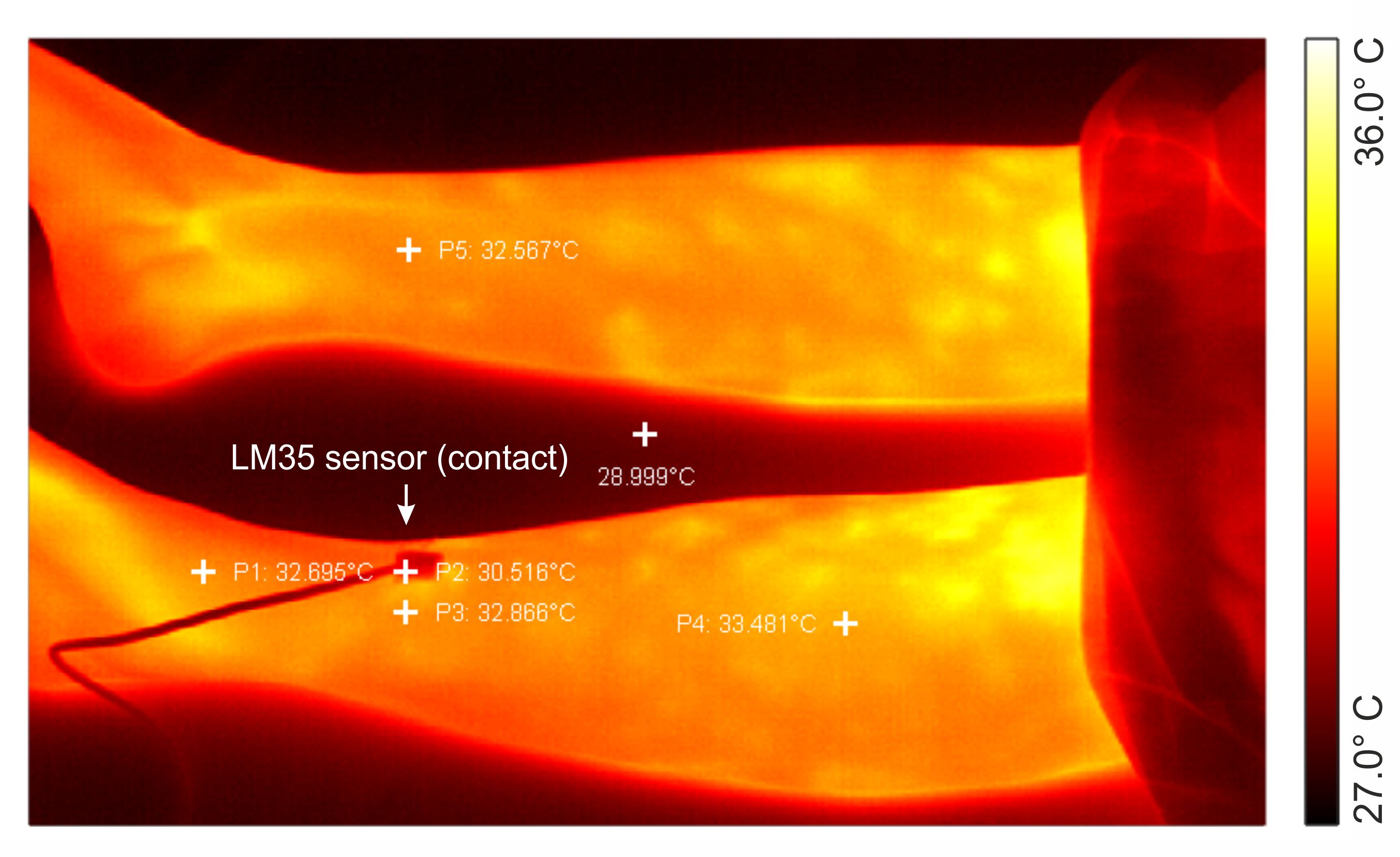}}
\caption{\small \textbf{(a)} The EEG, heart and breathing rate sensor for a real-time biofeedback and auditory control of ASC during meditation; \textbf{(b)} the LM35 precision temperature sensor fixed on the skin; \textbf{(c)}. IR thermal image, wavelength 8-14$\mu$m, NEDT $\leq60$mK (shown is the LM35 sensor measuring the skin surface temperature). \label{fig:Muse}}
\end{figure}
Meditators use the Muse-2 EEG, heart and breathing sensor, see Fig. \ref{fig:Muse}, for real-time auditory control of ASC. Experiments in \cite{KernbachOperator19nq} showed that the use of real-time biofeedback in ASC has a positive effect on 'active meditation' practices with long-term focused attention. Body temperature is measured in two ways: skin and core temperature by contact LM35 precision sensor, connected to a high resolution logger and thermal emission in 8-14$\mu m$ area by IR imaging sensor (NEDT $\leq60$mK). Both methods deliver the same temperature data, but differ significantly in terms of thermal dynamics: IR method requires open-skin regions that quickly cool down, while the contact method is best suited for measuring internal body dynamics. 

\begin{figure}[htp]
\centering
\subfigure{\includegraphics[width=0.45\textwidth]{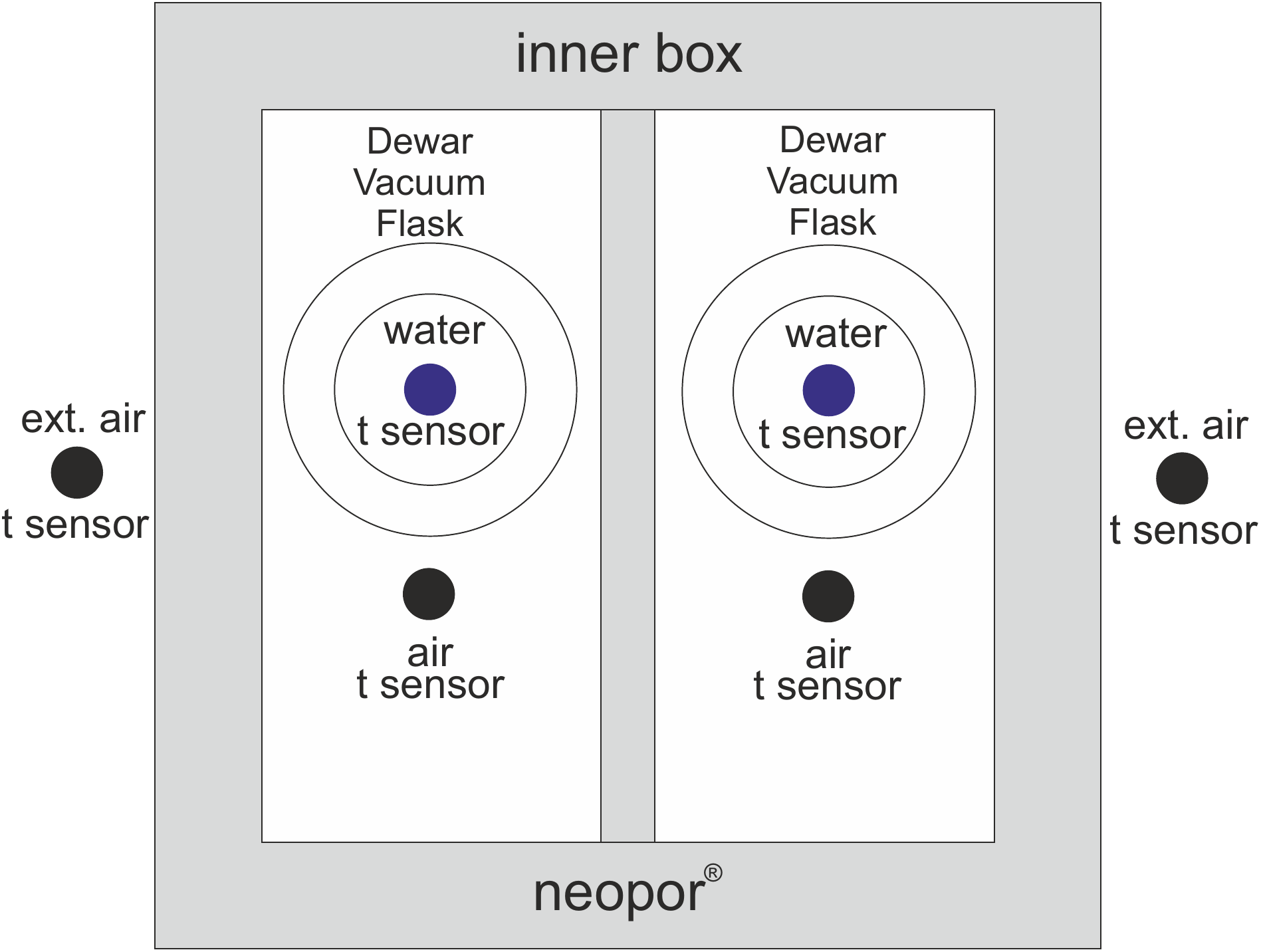}}
\subfigure{\includegraphics[width=0.49\textwidth]{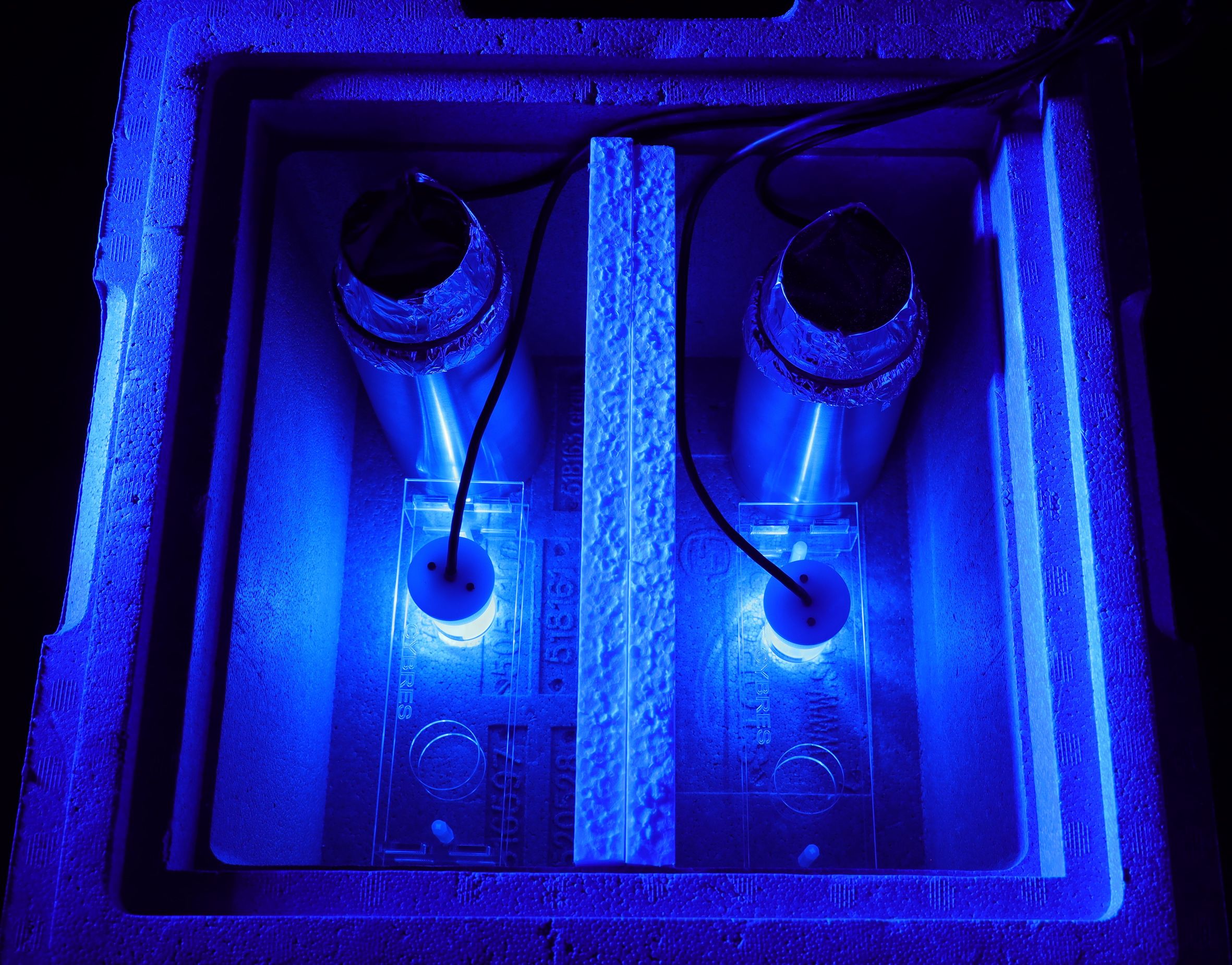}}
\caption{\small Passive differential calorimeter. \textbf{(a)} Scheme of the setup; \textbf{(b)} Image of the system with two Dewar vacuum flasks with NTC thermistor immersed in 15ml of water and two air temperature sensors.   \label{fig:calorimeter}}
\end{figure}

Setup with differential calorimeter is shown in Fig. \ref{fig:calorimeter} and extends the earlier work \cite{Gurtovoi92en}. One channel is used for experiments, the second channel represents a control sensor. Each channel has 15 ml of water in Dewar vacuum flask with NTC thermistor immersed in the fluid. Additionally, the air temperature is monitored in each section of the calorimeter inside and outside of the thermo-insulating box. Stabilized power supply is also measured by a separate sensor, the delta-sigma ADC with 24 bit sampling is used. In this work we utilize a passive scheme for the calorimeter, where a slow temperature variation is provided by a circadian rhythm in the laboratory (about 1-1.5 $^\circ$C for 24 hours) and guaranties a high degree of homogeneity in temperature distribution. Differences caused by changes in the heat capacity of water samples are measured as symmetry breaks between channels, small fluctuations about $10^{-5}$ $^\circ$C of relative temperature can be detected. This system is used to explore the heat capacity of para- and ortho- isomers of water \cite{kernbach23Thermal}. In several experiments we use electrochemical impedance spectrometers \cite{Kernbach17water} to detect effects on environmental sensors, see Fig. \ref{fig:monksData}.

Two separated rooms have been prepared for such experiments, see Fig. \ref{fig:plan}: all PCs and heat producing elements were removed from measurement laboratory; thermo-insulating screens have been installed in windows and doors; the room was closed and operated without human intervention.

\subsection{Methodology of experiments}

\begin{figure}[htp]
\centering
\subfigure[\label{fig:monksA}]{\includegraphics[width=0.4\textwidth]{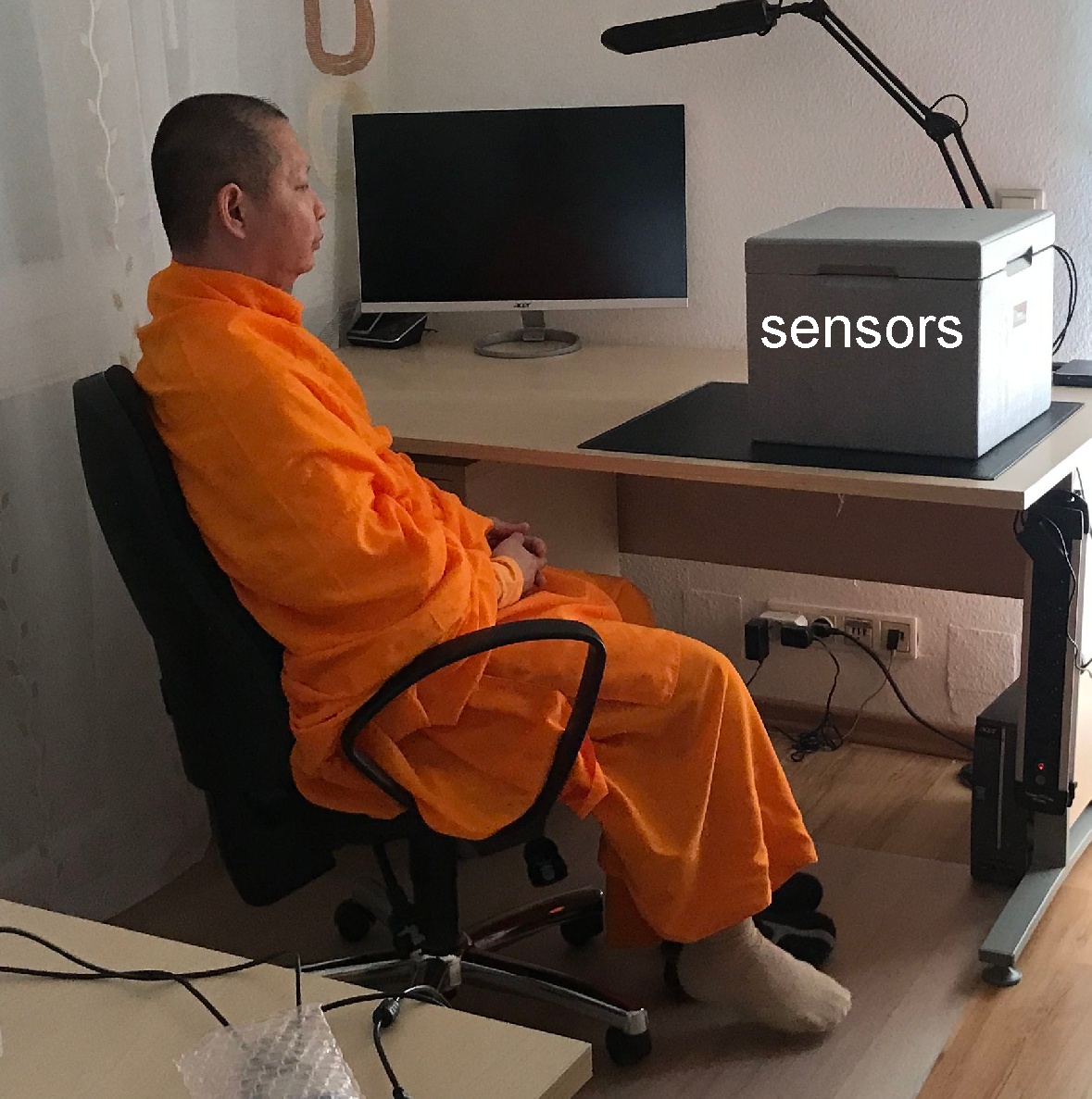}}
\subfigure[\label{fig:monksB}]{\includegraphics[width=0.4\textwidth]{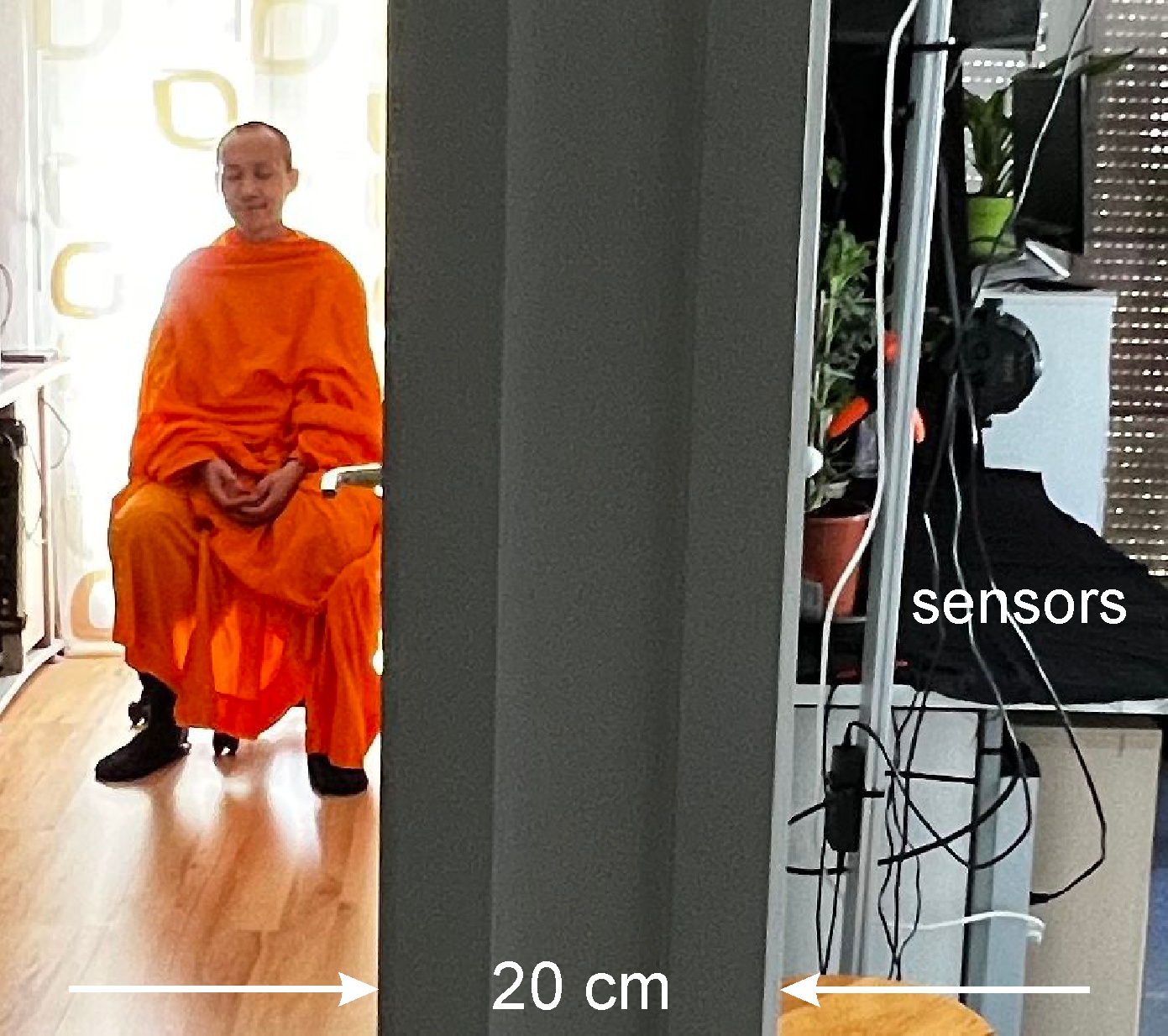}}
\subfigure[\label{fig:plan}]{\includegraphics[width=0.49\textwidth]{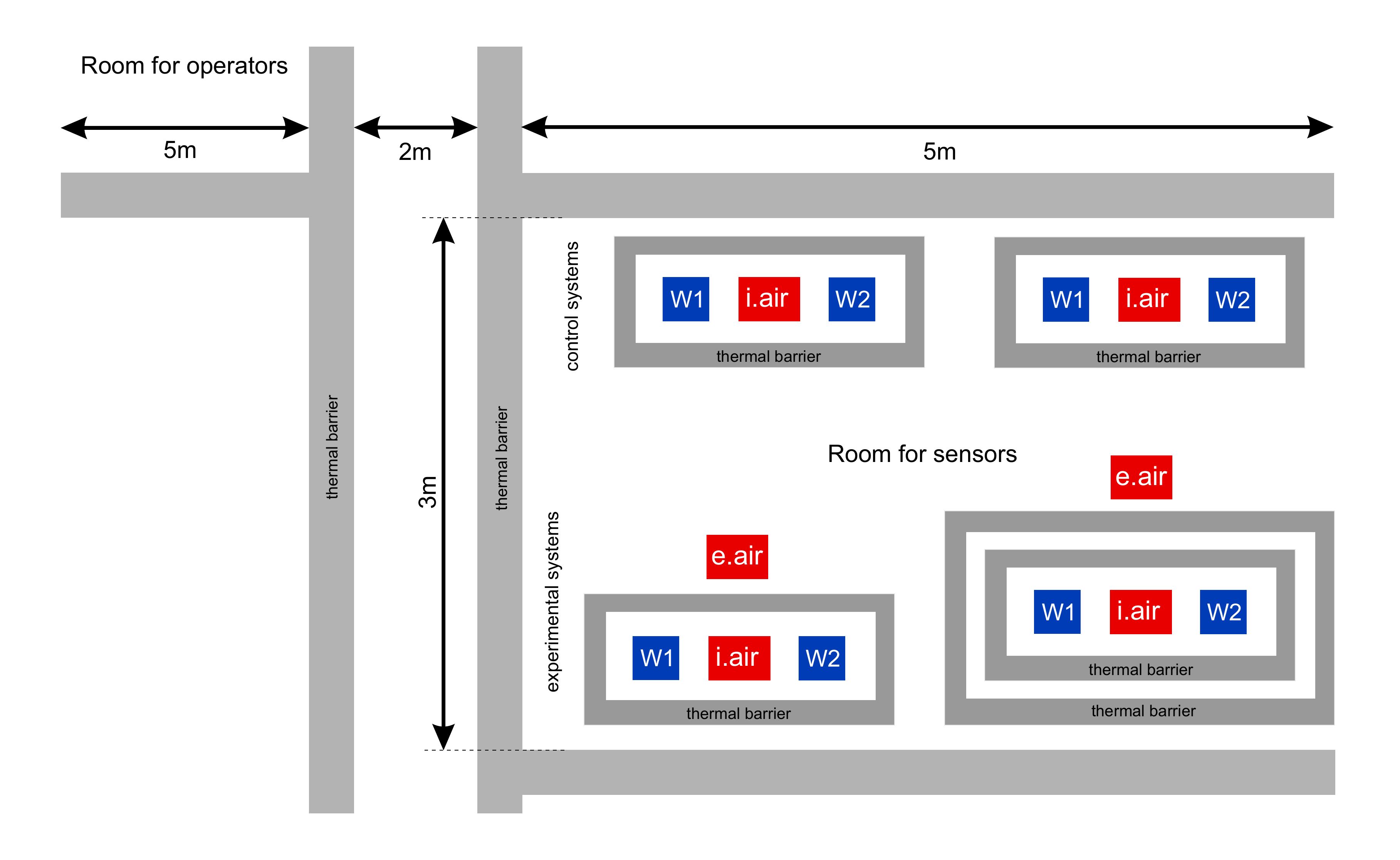}}
\caption{\small \textbf{(a,b)} The phase 1 of experiments on studying the impact of a meditator in ASC on close surrounding objects at a distance of 1-3 meters with and without obstacles. \textbf{(c)} Scheme of two laboratories for the phase 2 of experiments, W1, W2 -- water sensors, i.air, e.air -- internal and external air sensors, totally 3 experimental and 3 control differential calorimeters are used experiments. \label{fig:monks}}
\end{figure}

Experiments are divided into two phases, with attention focused inside (the phase 1) and outside (the phase 2) the body. Experiments in the phase 1 are separated into three subgroups 1a, 1b, 1c and are also aimed at studying the effect of a meditator in ASC on closely spaced sensors. Since the human person irradiates heat, produces electrostatic fields, mechanical impacts and other 'classical interactions', measures were taken to isolate 'classical' from 'non-classical' factors. For example, a heat-insulating container with 3-5 liters of water delays the penetration of heat inside for 20-25 minutes, see Fig. \ref{fig:monksA}; this enables conducting short-term meditative experiments. In other cases, the meditators were separated from the measurement systems by a reinforced concrete wall, see Fig. \ref{fig:monksB}.

Experiments in the phase 2 are conducted with meditators in the operator room and differential calorimeters installed in measurement room (the distance is about 7-10 meters), separated by two concrete walls, see Fig. \ref{fig:plan}. A heat transfer between rooms takes about 3-4 hours. Thus, we can exclude a direct thermal influence of meditators on calorimetric sensors during experiments in the phase 2.

\begin{figure}[htp]
\centering
\subfigure{\includegraphics[width=0.49\textwidth]{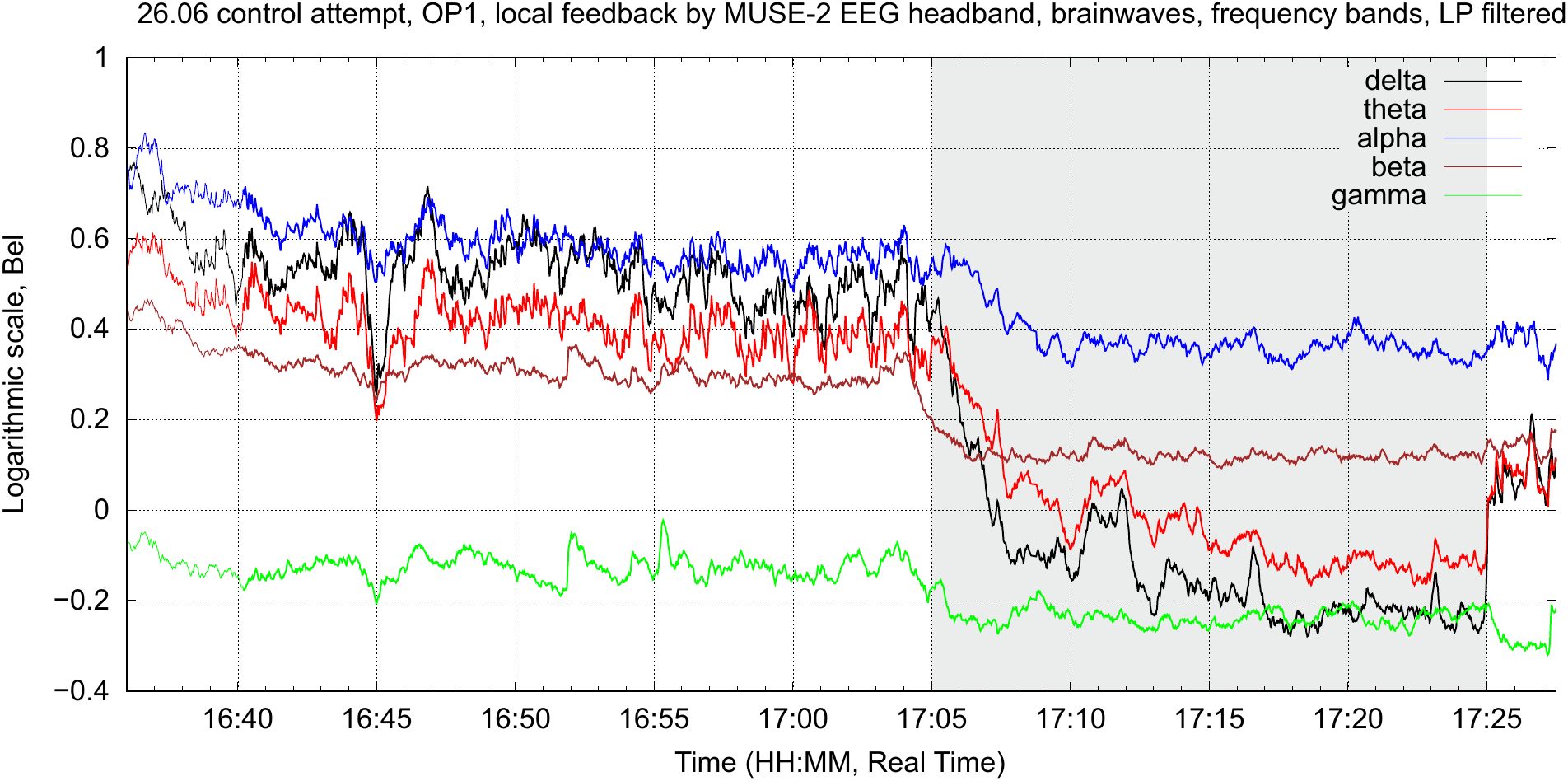}}~
\caption{\small Example of EEG data in test experiments without biofeedback. The ASC area is shown as a gray bar. \label{fig:EEGExample}}
\end{figure}

The meditative visualization during all experimental phases follows the tantric Tibetan techniques \cite{Mullin97,Tsongkhapa14} and Dhammakaya meditation \cite{SUDSUANG1991543}. The key parameters are long-term focused attention in ASC on a physical object (or internal organs in the body) and the state of consciousness (its cognitive and behavioral patterns in ASC). Example of EEG data in test experiments is shown in Fig. \ref{fig:EEGExample}.

As in other ASC experiments \cite{Kozhevnikov13, Brechet22}, two-stage EEG dynamics with different microstates is observed. It should be noted that the initial stage of ASC takes about 10-15 minutes, when cognitive functions are restructuring from processing of external signals to internal visualizations. The depth of ASC depends on external distorting stimuli, mental efforts in maintaining the focus of attention, psychic and muscle relaxation.

Evaluation of results follows the single factor methodology from other studies of Buddhist techniques \cite{Kozhevnikov13, SUDSUANG1991543, Amihai14}. The results are anonymized and randomized as much as possible for a small group (5 practitioners in age from 25 to 50 years with meditation experience ranged from 5 to 10 years) to remove bias.

\section{Experiments}

An overview of all attempts and their parameters is given in Table \ref{tab:parameters}. Experiments in the phase 1a and 1b are fully reproducible in the selected group of meditators, we observe only a variation in intensity of temperature responses. We also observed a non-temperature response of surrounding electrochemical sensors on presence of meditators in all cases, however, detecting the exact begin and end of exercises, as shown in Fig. \ref{fig:monksData}, was possible only in several attempts. Since this topic represents the focus of other works \cite{KernbachOperator19nq,KernbachQigong22en,KernbachMahirishi20en}, it was not pursued further here. Control measurements were carried out a) in parallel to experimental sessions on separate calorimeters (as non-targeted sensors); b) during so-called 'empty sessions' (meditators were present but no sessions were conducted), several examples of control attempts are shown in Figs. \ref{fig:phase2ExpControl1}-\ref{fig:phase2ExpControl3}. 

\begin{table}[h]
\begin{center}
\caption{\small Parameters of performed experiments, 'N of sensors' -- different sensors used in one attempt. \label{tab:parameters}}
\fontsize {9} {10} \selectfont
\begin{tabular}{
p{0.7cm}@{\extracolsep{3mm}}
p{1.1cm}@{\extracolsep{3mm}}
p{1.1cm}@{\extracolsep{3mm}}
p{1.1cm}@{\extracolsep{3mm}}
p{3.5cm}@{\extracolsep{3mm}}
}\hline \hline
phase &N attempts & N sensors & N \mbox{positive} & description \\\hline
1a     & 12    & 1     & 12     &  body temperature \\
1b     & 10    & 4     & 10     &  hand temperature \\
1c     & 7    & 5     & --      &  surrounding objects \\
2		   & 63   & 18(3x6)         & 57    &  external calorimeters \\
2		   & 67   & 18(3x6)         & 11    &  control (most positive responses in non-targeted sensors)  \\\hline
       &&& 0.00012 & Chi-square test, p value \\  
		   &&& 0.0008  & U test, p value\\\hline
total  & 159  & \multicolumn{3}{l}{independent experiments}\\
       & 2427 & \multicolumn{3}{l}{operator-sensor sessions}\\
\hline\hline
\end{tabular}
\end{center}
\end{table}

\subsection{The phase 1 experiments}

\textbf{1a. Body temperature effects} have been measured with three methods (IR, skin surface, core temperature). 
\begin{figure}[ht]
\centering
\subfigure[]{\includegraphics[width=0.49\textwidth]{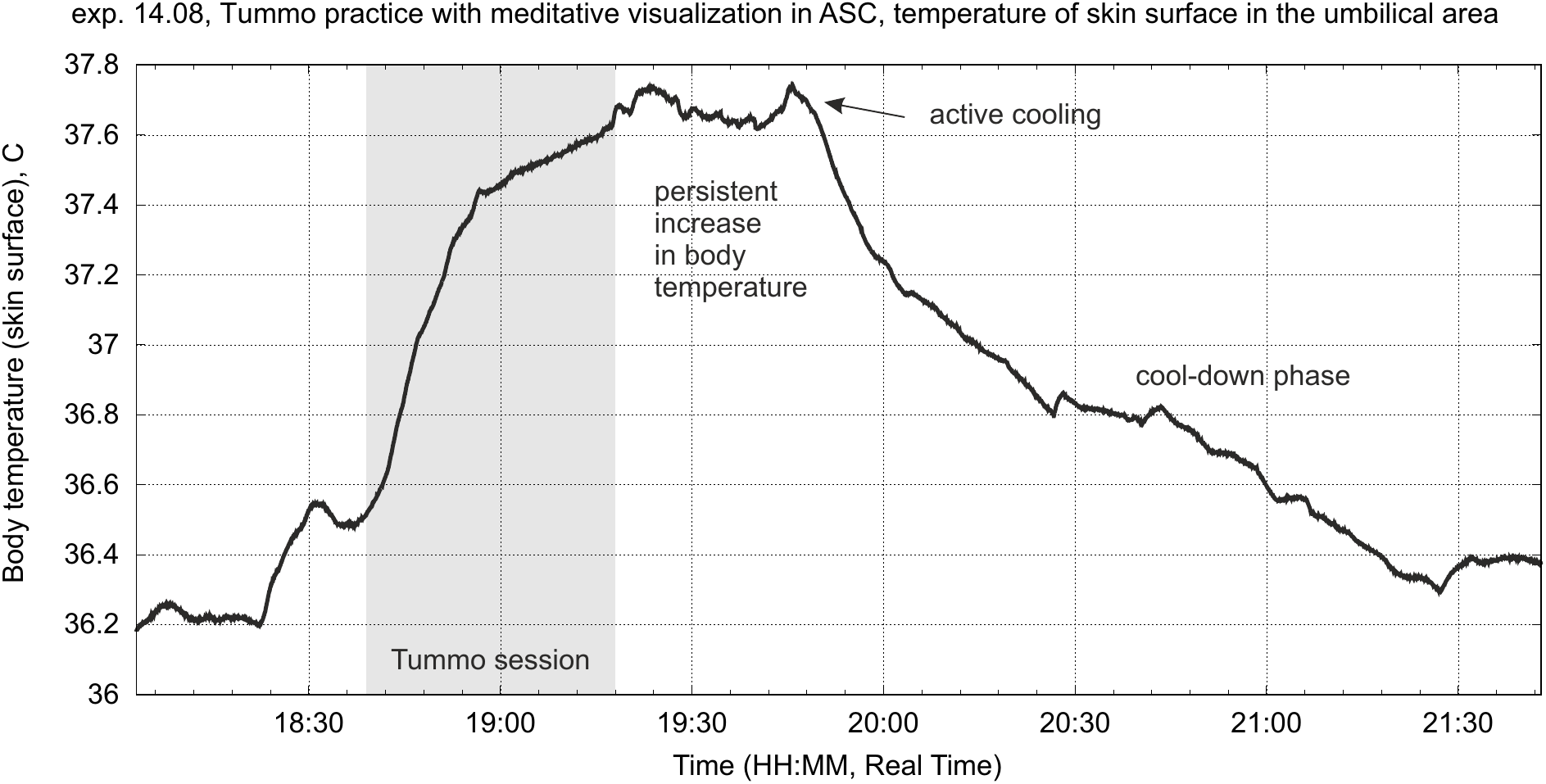}}
\subfigure[]{\includegraphics[width=0.49\textwidth]{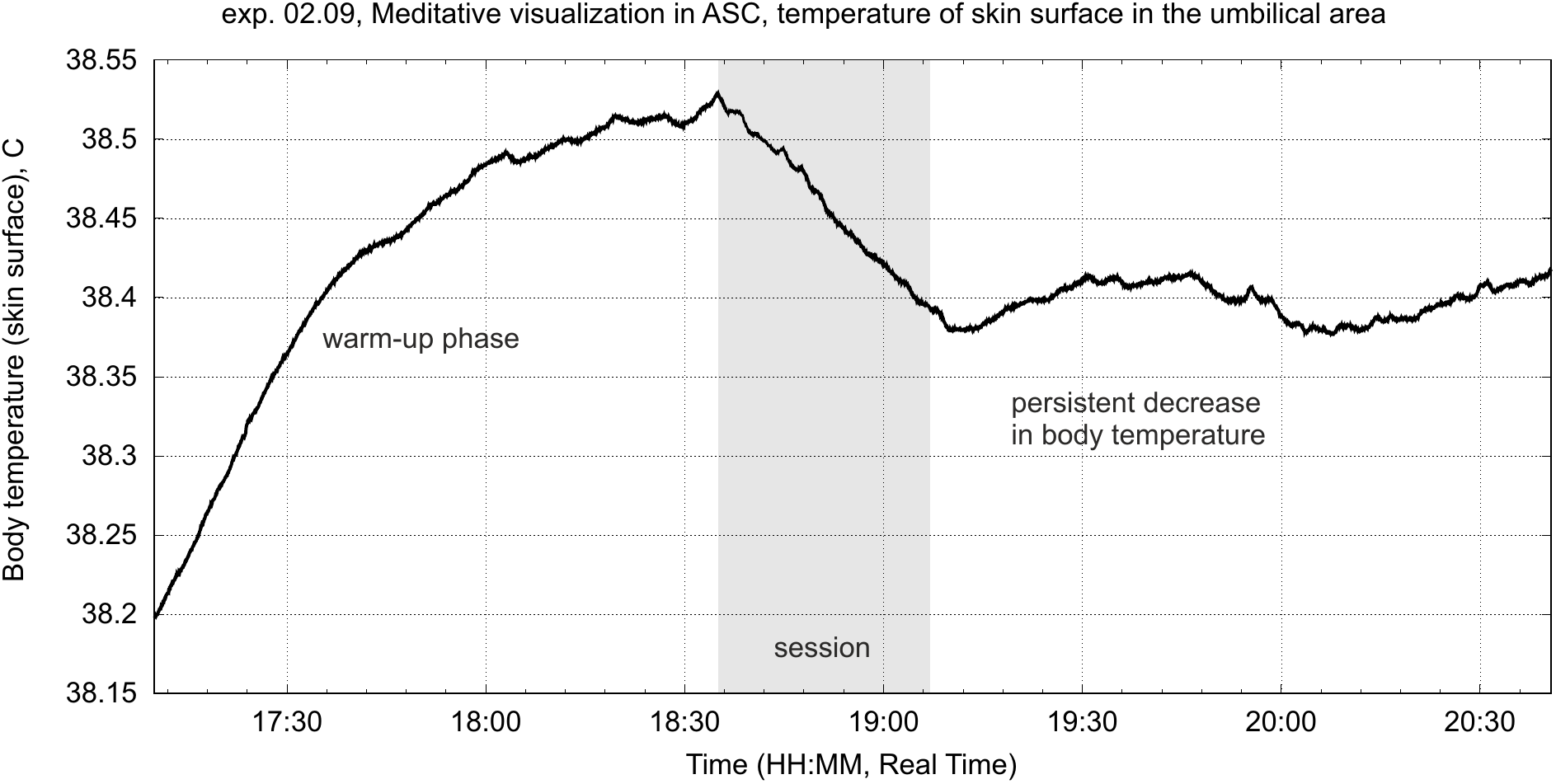}}
\subfigure[]{\includegraphics[width=0.49\textwidth]{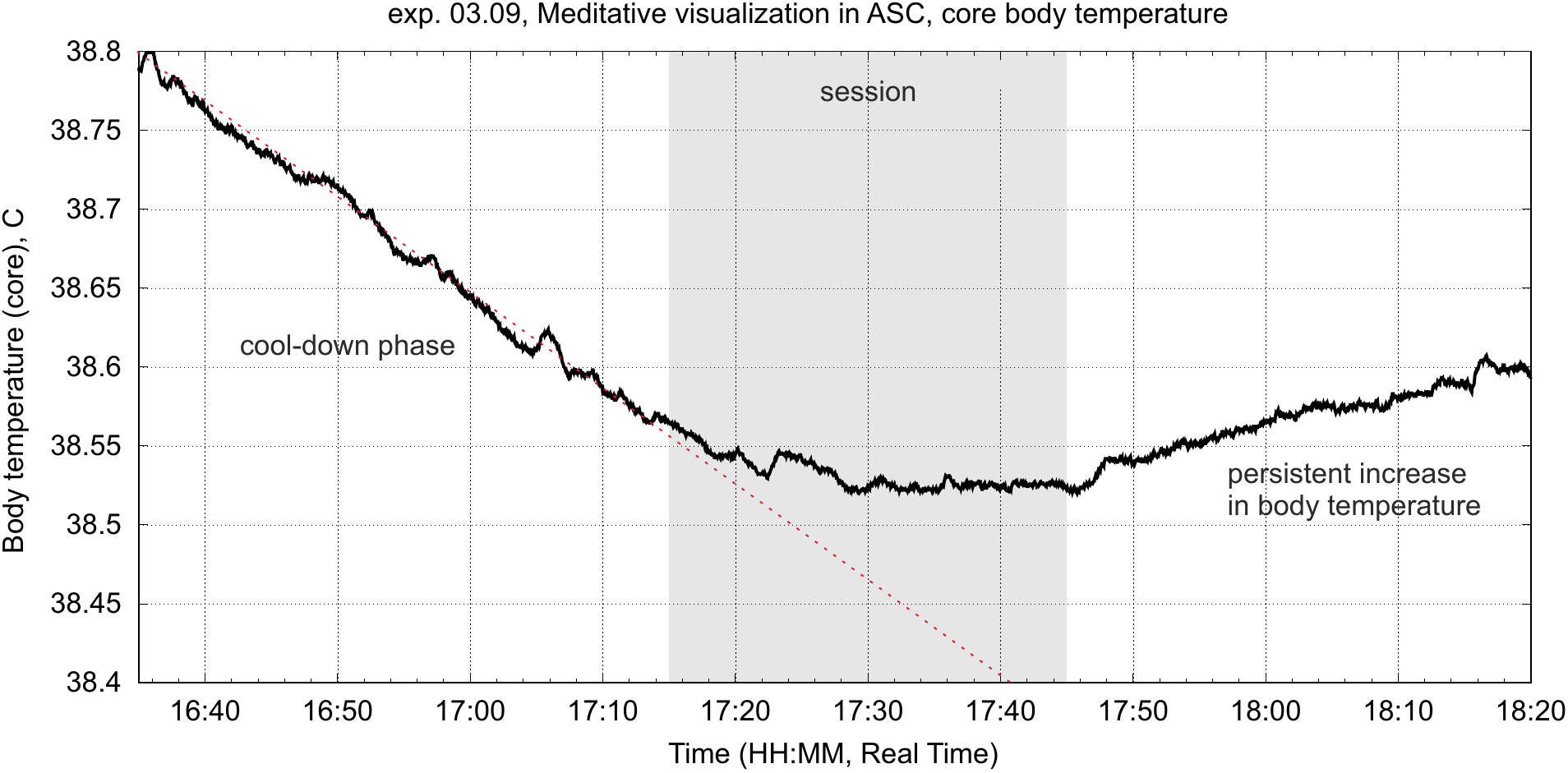}}
\caption{\small Examples of typical temperature effects for meditative visualization in ASC: \textbf{(a)} Tummo session with increasing the body temperature (measured by contact skin surface method in abdominal region near the navel); \textbf{(b)} Using meditative visualization for lowering body temperature (measured by contact skin surface method in abdominal region near the navel); \textbf{(c)} Variation of core temperature during meditative visualization without breading exercises. Regions of persistent changes after the sessions is well visible in all cases. \label{fig:TummoTemp}}
\end{figure}
\begin{figure}[ht]
\centering
\subfigure{\includegraphics[width=0.49\textwidth]{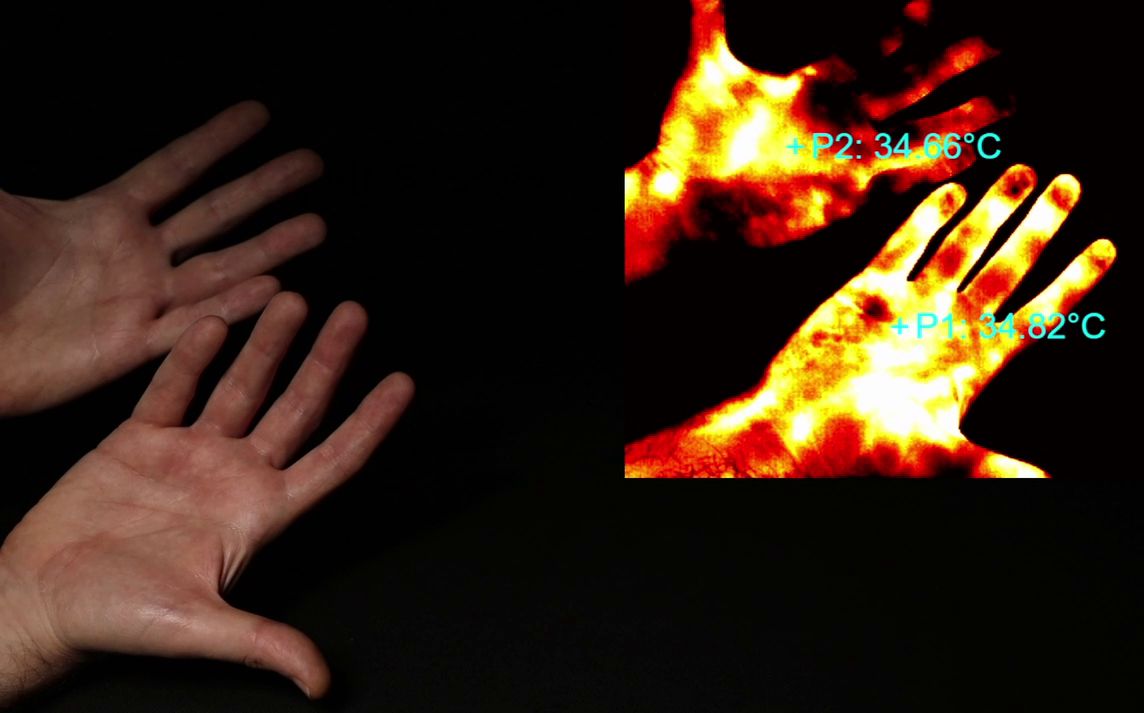}}
\subfigure[]{\includegraphics[width=0.49\textwidth]{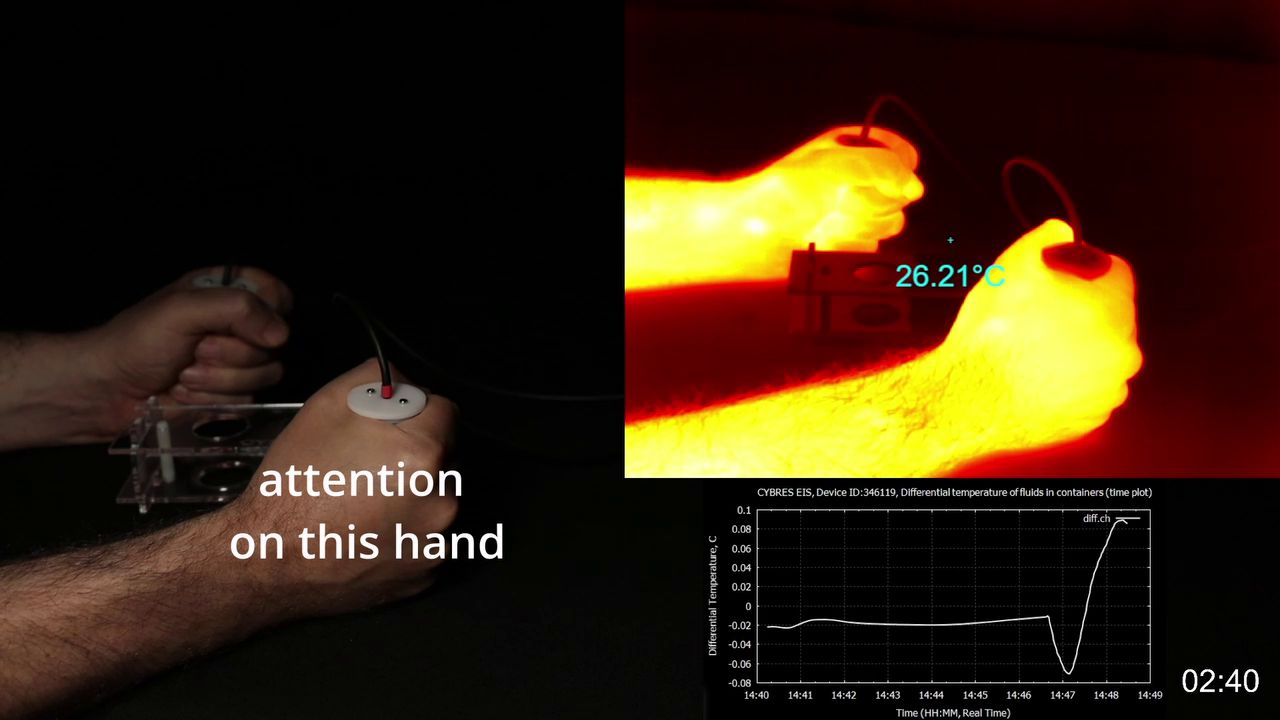}}
\subfigure[]{\includegraphics[width=0.49\textwidth]{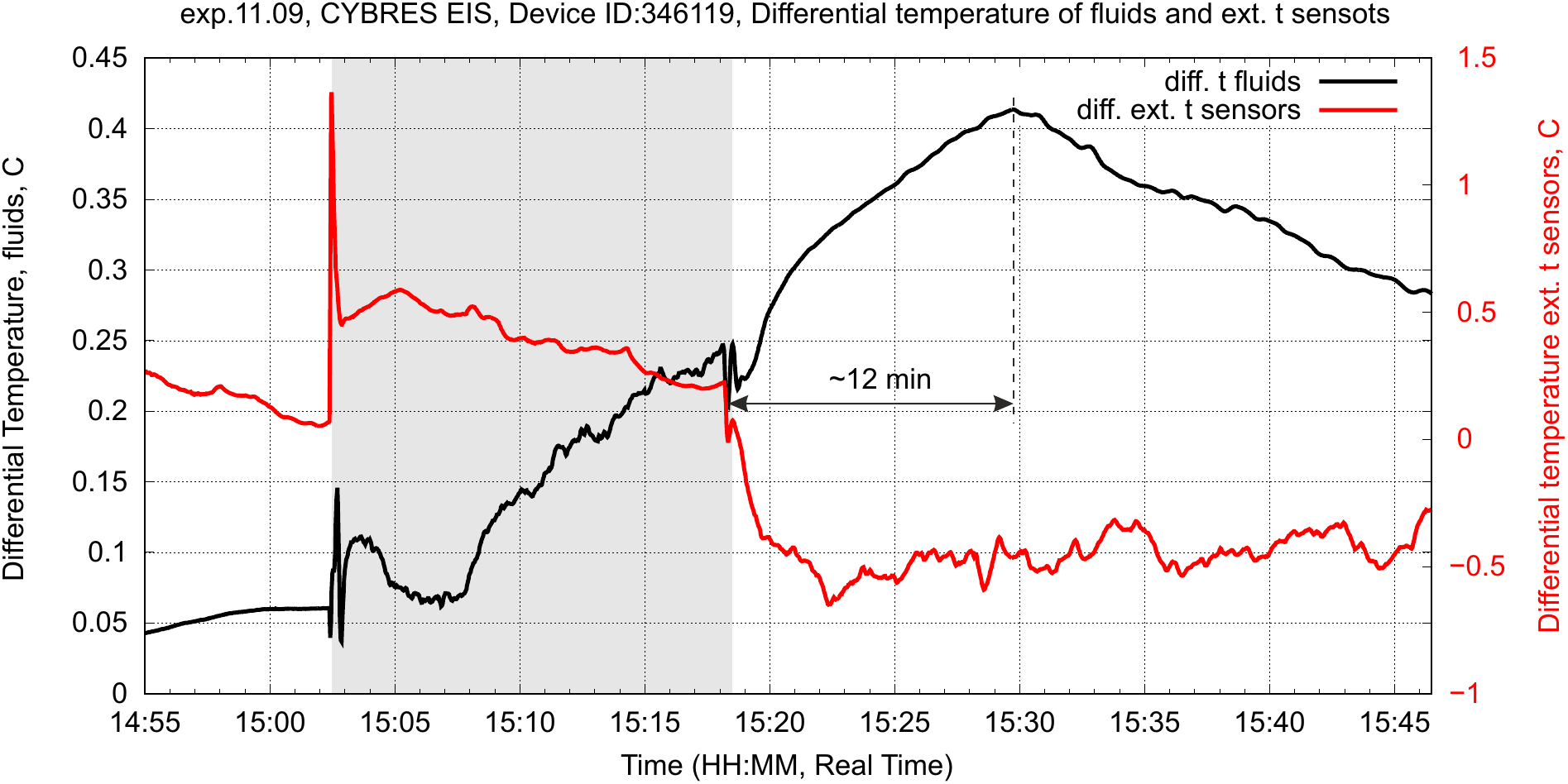}}
\caption{\small Calorimetric measurement (heating up 15 ml water in small containers) with focusing attention on one hand, each container has one \textit{t} sensor immersed in the fluid and one \textit{t} sensor measuring air temperature outside the container: \textbf{(a)} Setup; \textbf{(b)} Persistent temperature dynamics after removing the water containers from the hands lasts about 12 minutes, the gray area indicates the time the containers were held in hands. \label{fig:HandsTemp}}
\end{figure}
The temperature data depends essentially on behavior of a meditator, even small changes in position (during a 2-3 hour experiment) disturb a thermodynamic equilibrium of the body and are reflected in an increasing or decreasing trend of \textit{t}. Temperature dynamics depends on a fat content of tissues, clothing, room temperature and other factors that affect thermodynamic conditions. Therefore, experiments have been performed in a fixed position of a meditator, providing enough time for stabilizing body temperature before and after attempts. Fig. \ref{fig:TummoTemp} demonstrates several experiments with raising or lowering body temperature through meditative visualization in ASC. We observe an almost immediate response of body temperature to the visualization, the area of persistent changes after the session is clearly visible; its duration is longer than the session itself. Inclination of temperature dynamics during the sessions is higher for regions of 'natural' warm-up and cool-down phases, this indicates their more intensive thermodynamic mechanisms. Experiments on lowering body temperature does not include breading exercises.

\textbf{1b. Additional attempts with heating up 15 ml water in small containers} have been undertaken to measure the heat produced by hands if focusing attention on one of the hands. Each container has one \textit{t} sensor immersed in the fluid and one \textit{t} sensor measuring air temperature outside the container. Without attention, hands produce almost equal amount of heat that results in a stable differential temperature of water and air sensors. Focusing attention on one hand increases its skin surface temperature up to 1.5-2 $^\circ$C and increase the differential temperature of water. We also observe persistent changes after the water containers are removed from hands for about 12 minutes -- increasing temperature of one fluid but equal temperature of air outside, see Fig. \ref{fig:HandsTemp}. Such a dynamics is difficult to explain by heating only, and motivated a development of external differential calorimeters and methodology of experiments in the phase 2.

\textbf{1c. Effects on surrounding electrochemical sensors} have been investigated multiple times in different laboratories \cite{Matos17,Gurtovoi92en,KernbachOperator19nq}, where cases of focused and unfocused attention are separated. Even in the unfocused case (e.g. only with specific expectations), the experimenter demonstrated evident bias in experimental data \cite{Korenbaum21}; the focused attention produces more significant changes. The results of the phase 1c are similar to such studies, regardless of separating barriers. 
\begin{figure}[ht]
\centering
\subfigure{\includegraphics[width=0.49\textwidth]{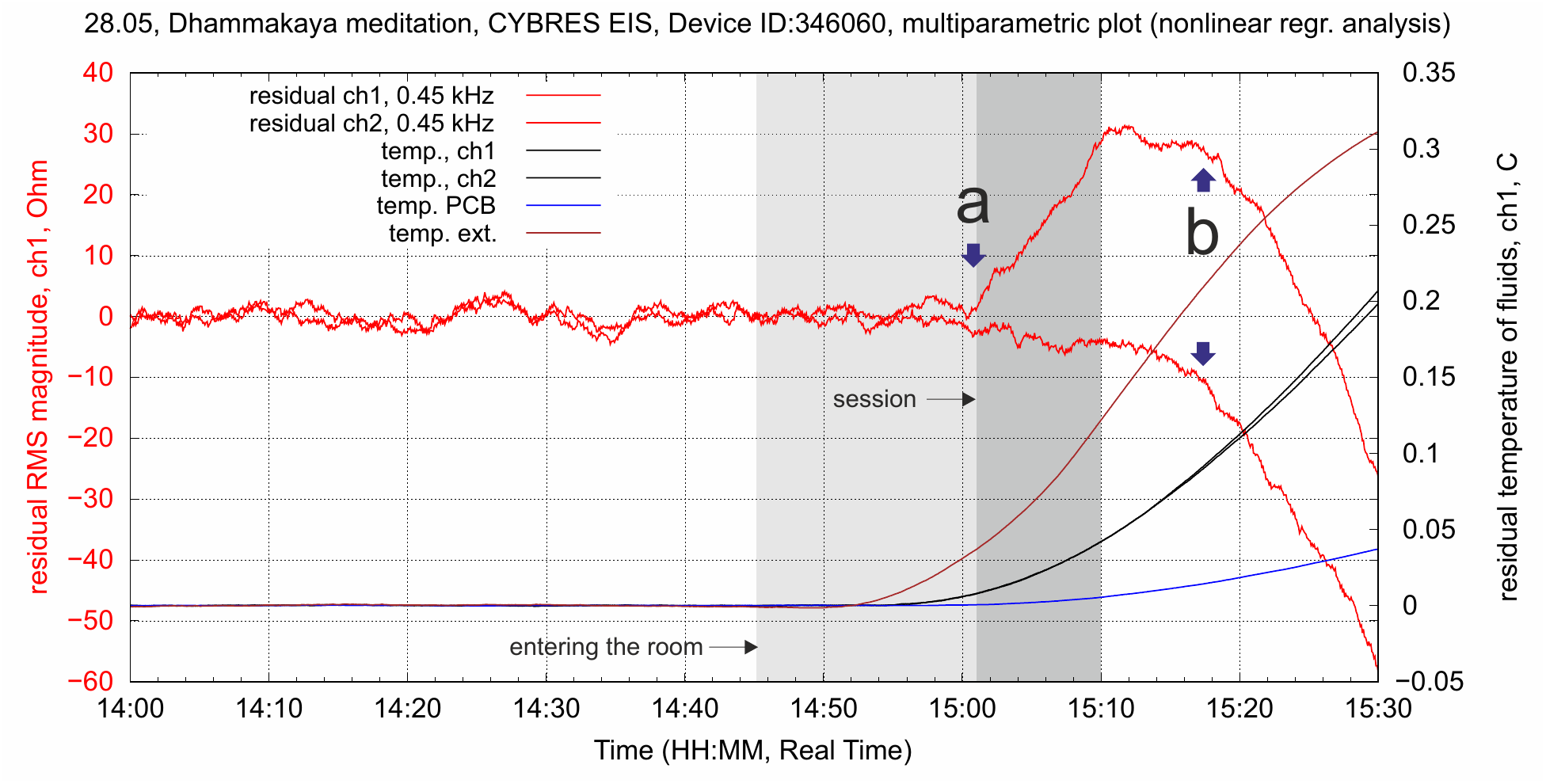}}
\caption{\small Effects on surrounding electrochemical sensors in the phase 1c: the point 'a' -- enter in the ASC (begin and end of ASC is clearly visible), the point 'b' -- thermal effects, generated by a body heat of meditator. The point 'a' is about 20 minutes before the point 'b'. \label{fig:monksData}}
\end{figure}

Example of electrochemical and thermal dynamics is shown in Fig. \ref{fig:monksData}. Enter in ASC is indicated by a rapid increase of ionic dynamics, see the point 'a'. Electrochemical dynamics follows the begin and end of meditation. As the meditator approaches the sensors, we observe a slow increase of temperature in the container. However, it starts to affect the electrochemical dynamics about 30-40 minutes later, see point 'b' in Fig. \ref{fig:monksData}. Thus, the dynamics between 'a' and 'b' points provides information about 'non-classing' effects of focused attention on the environment before 'classical' factors start affecting the sensor.

\subsection{The phase 2 experiments}
\label{sec:ExpPhase2}

The meditative visualization in this phase is similar to 'exteriorization of self', which is a basic technique of Vajrayana and other Buddhist (e.g. from Dzogchen lineage \cite{rinpoche2006union}) and Taoist \cite{chao1973taoist} traditions, and also explored in academic publications \cite{Menon16}. The meditator in the operator room attempts to visualize a predetermined channel of differential calorimeter placed in the measurement laboratory. During experiment, the meditator receives a real-time biofeedback (mostly acoustic signals based on EEG and breading rate) to control ASC and also real-time data from temperature sensors in the remote setup. 

\begin{figure}[htp]
\centering
\subfigure[\label{fig:typicalCalorimeterA}]{\includegraphics[width=0.49\textwidth]{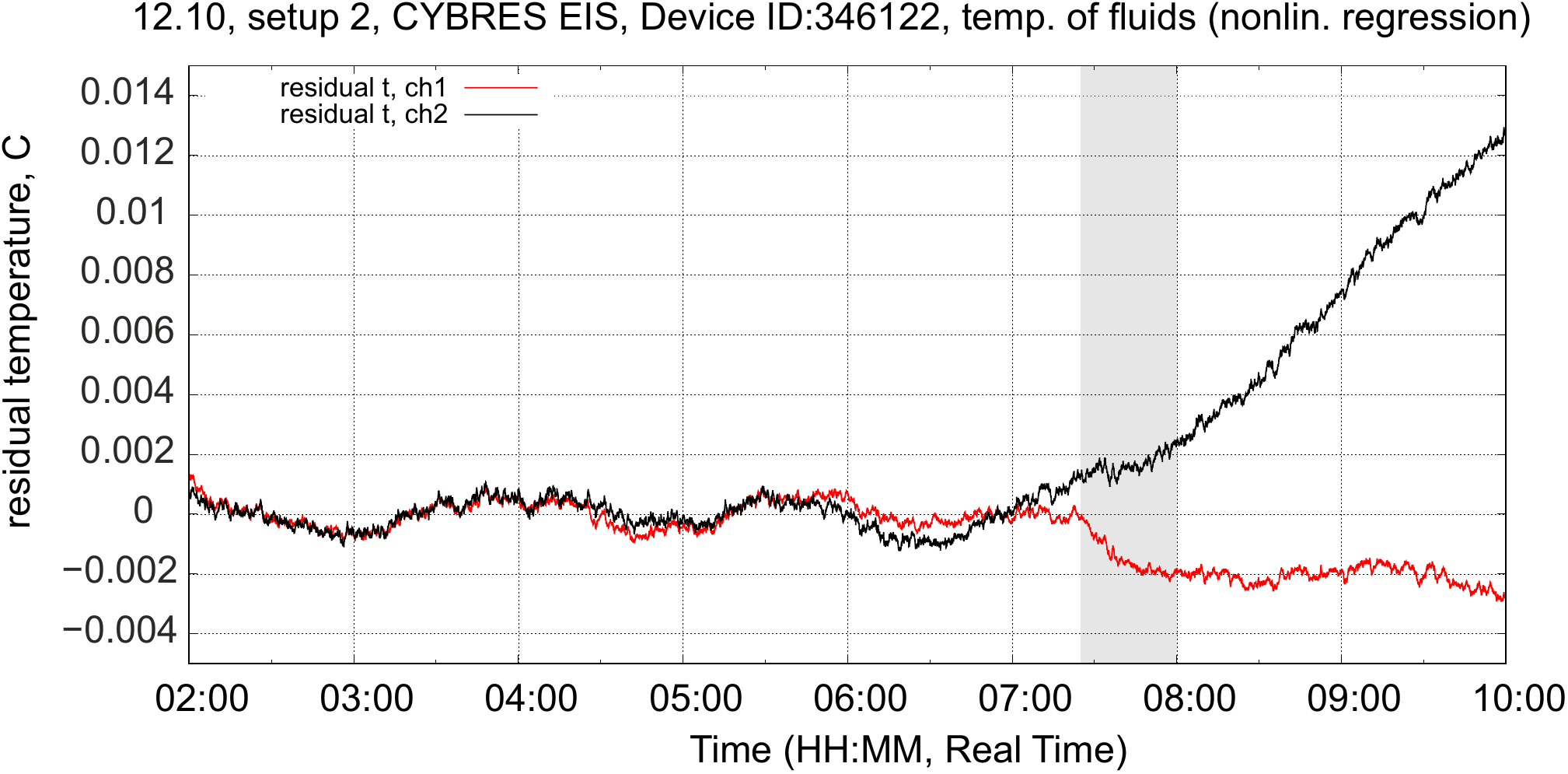}}
\subfigure[]{\includegraphics[width=0.49\textwidth]{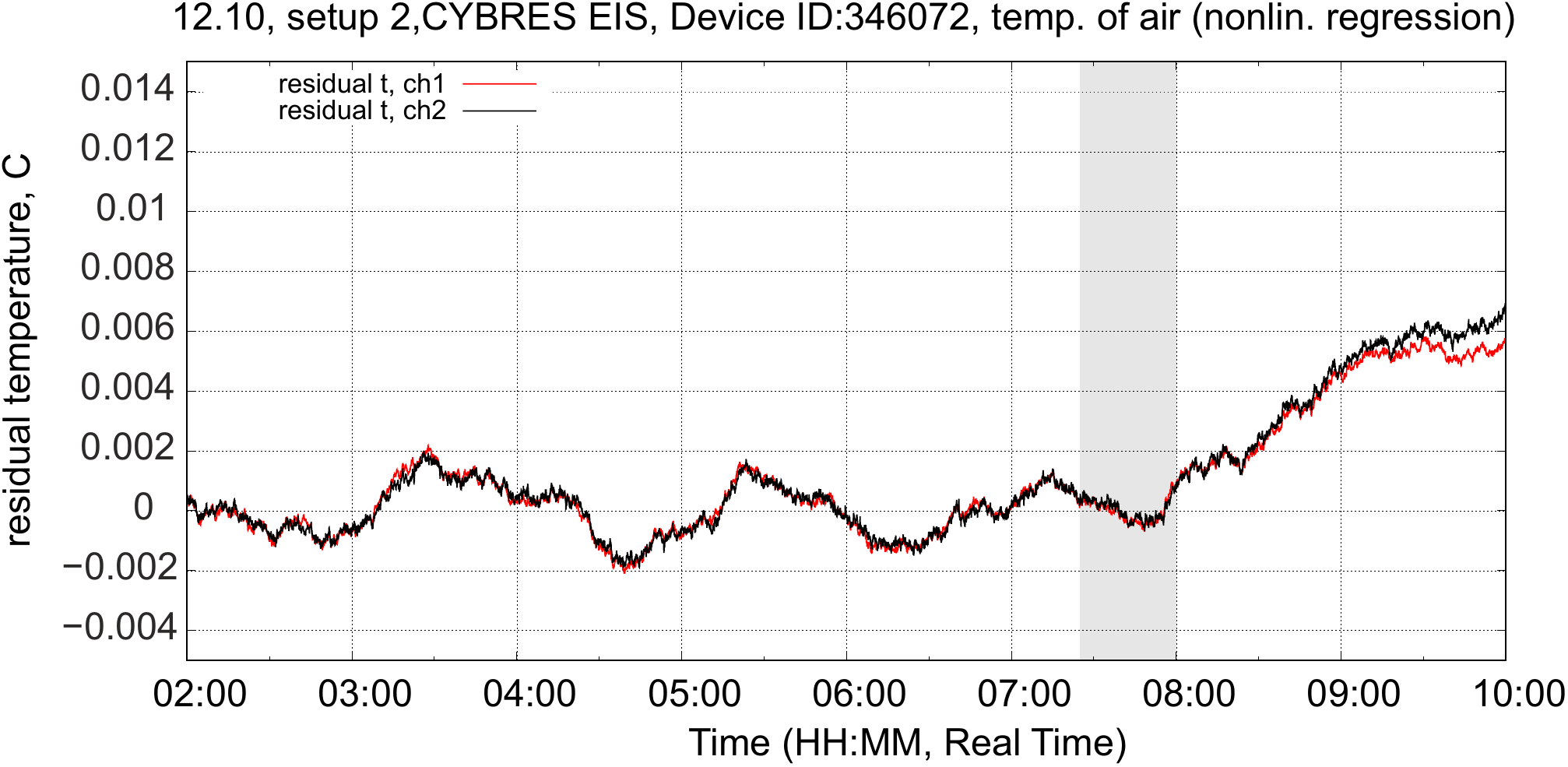}}
\caption{\small Example of experiments in the phase 2: \textbf{(a)} symmetry breaking of temperature in fluid sensors; \textbf{(b)} air sensors demonstrate a similar temperature of both channels. Grey bar shows an experimental session. These two graphs can be represented as differential temperatures in one plot, see Fig. \ref{fig:phase2ExpTwo}. \label{fig:typicalCalorimeter}}
\end{figure}

Typical dynamics of calorimeter data is shown in Fig. \ref{fig:typicalCalorimeter}. The targeted channel demonstrates a temperature deviation that results in symmetry breaking dynamics of fluid sensors, air sensors do not demonstrate any similar effects (i.e. this points to internal mechanisms in aqueous solutions that cause such thermal fluctuations). This dynamics is similar to the measurement in Fig. \ref{fig:HandsTemp}, but in this case a direct heat transfer from the mediator is excluded.

\begin{figure}[htp]
\centering
\subfigure[]{\includegraphics[width=0.49\textwidth]{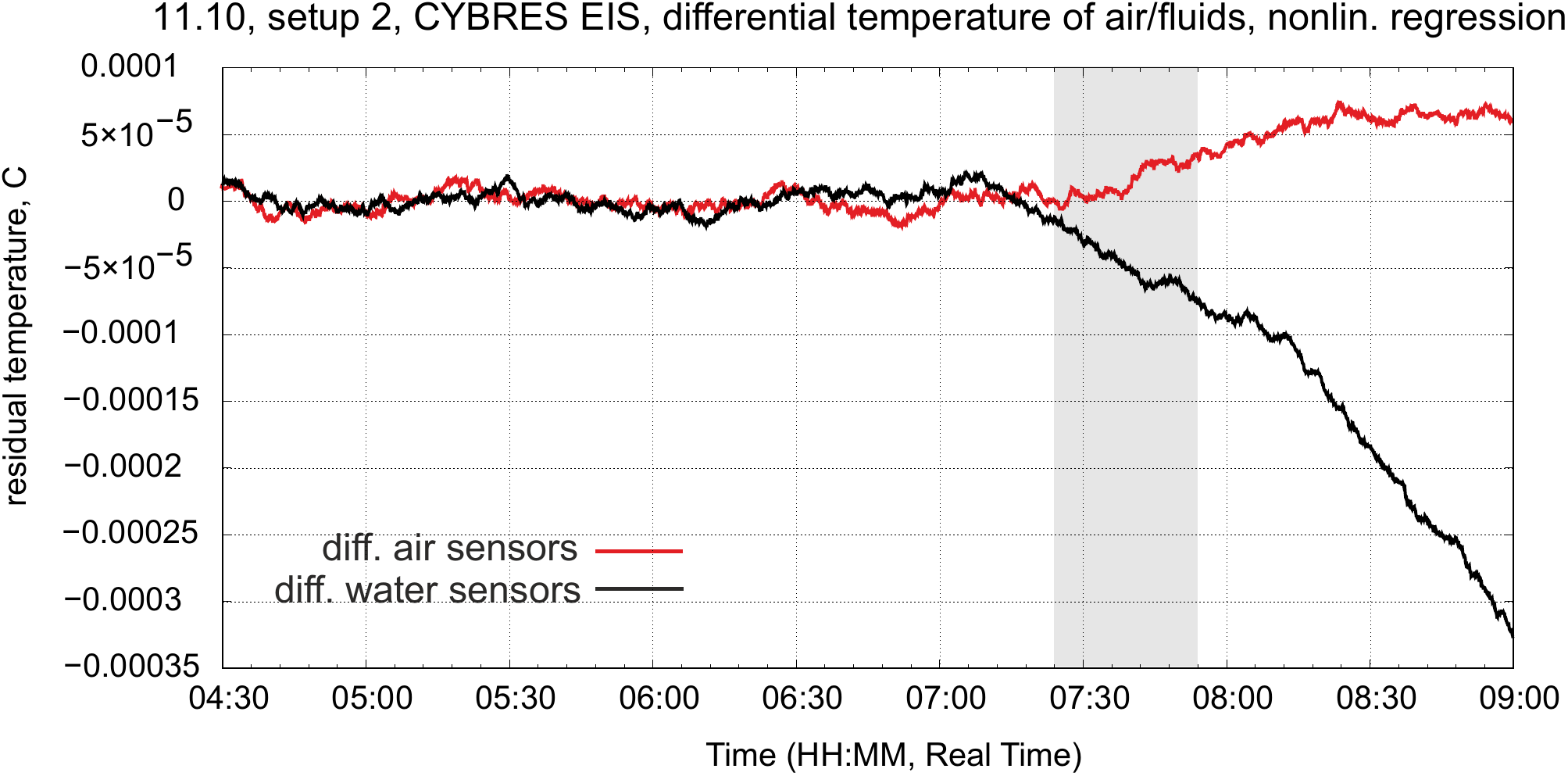}}
\subfigure[]{\includegraphics[width=0.49\textwidth]{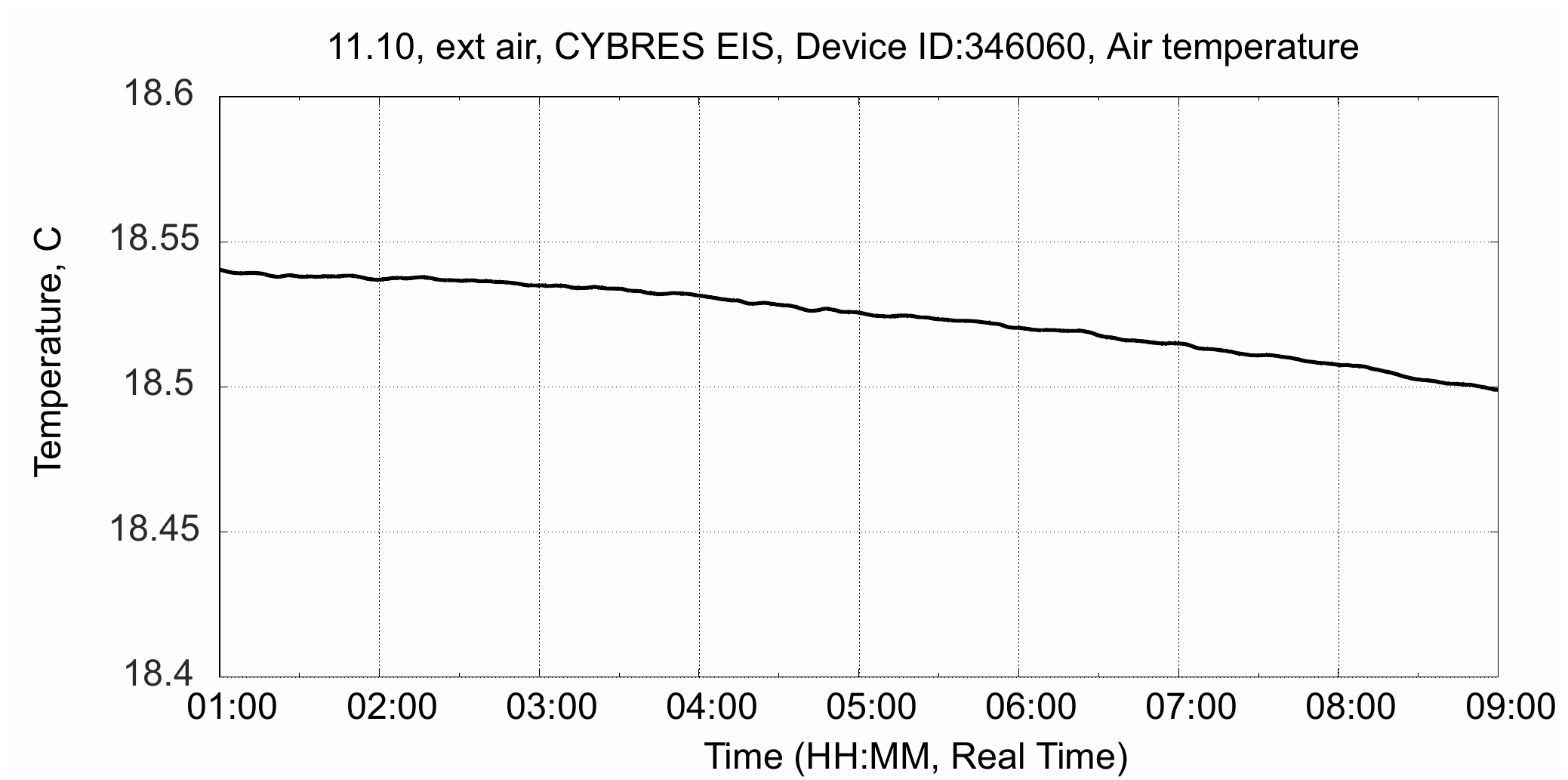}}
\subfigure[]{\includegraphics[width=0.49\textwidth]{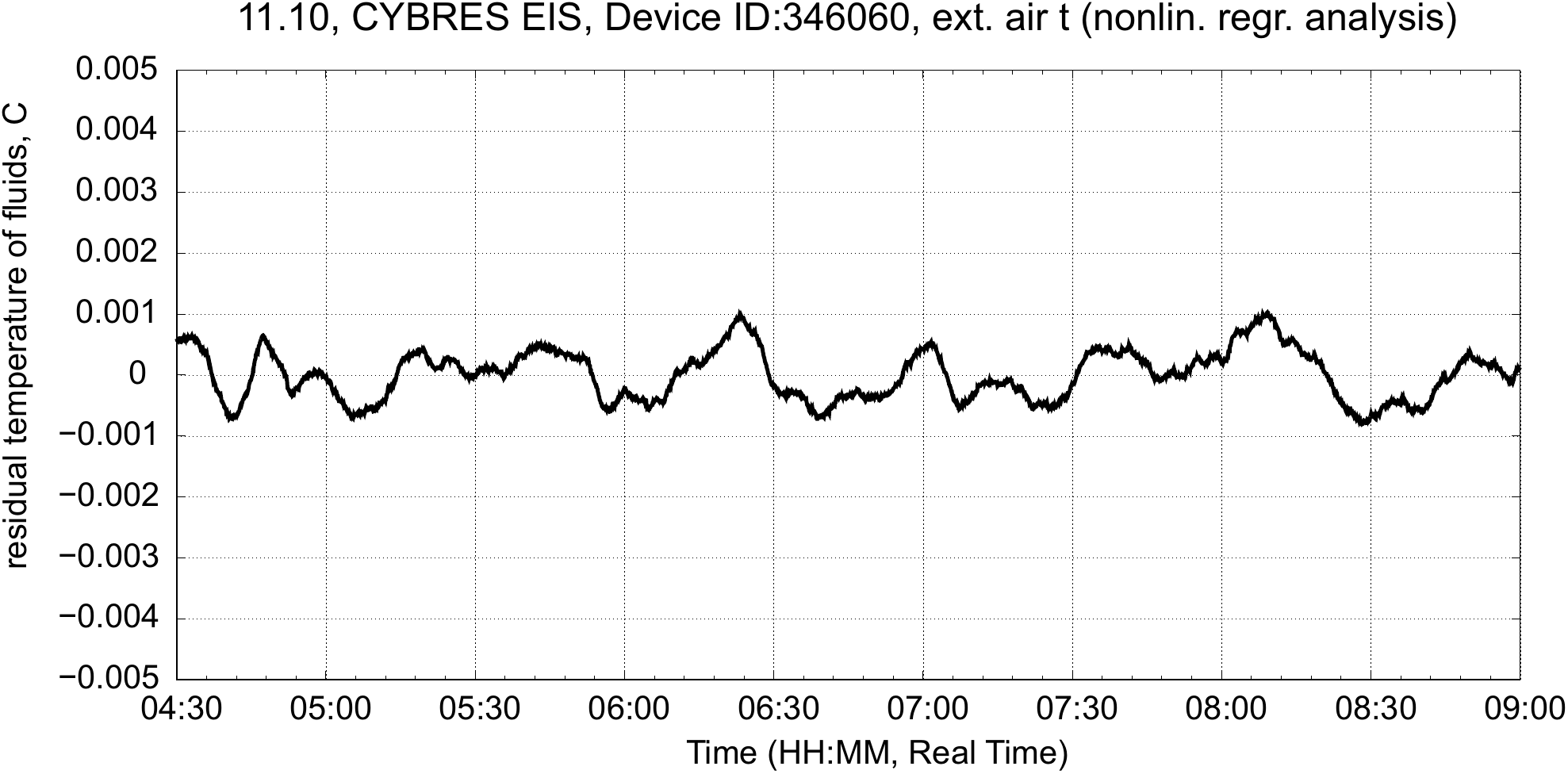}}
\subfigure[]{\includegraphics[width=0.49\textwidth]{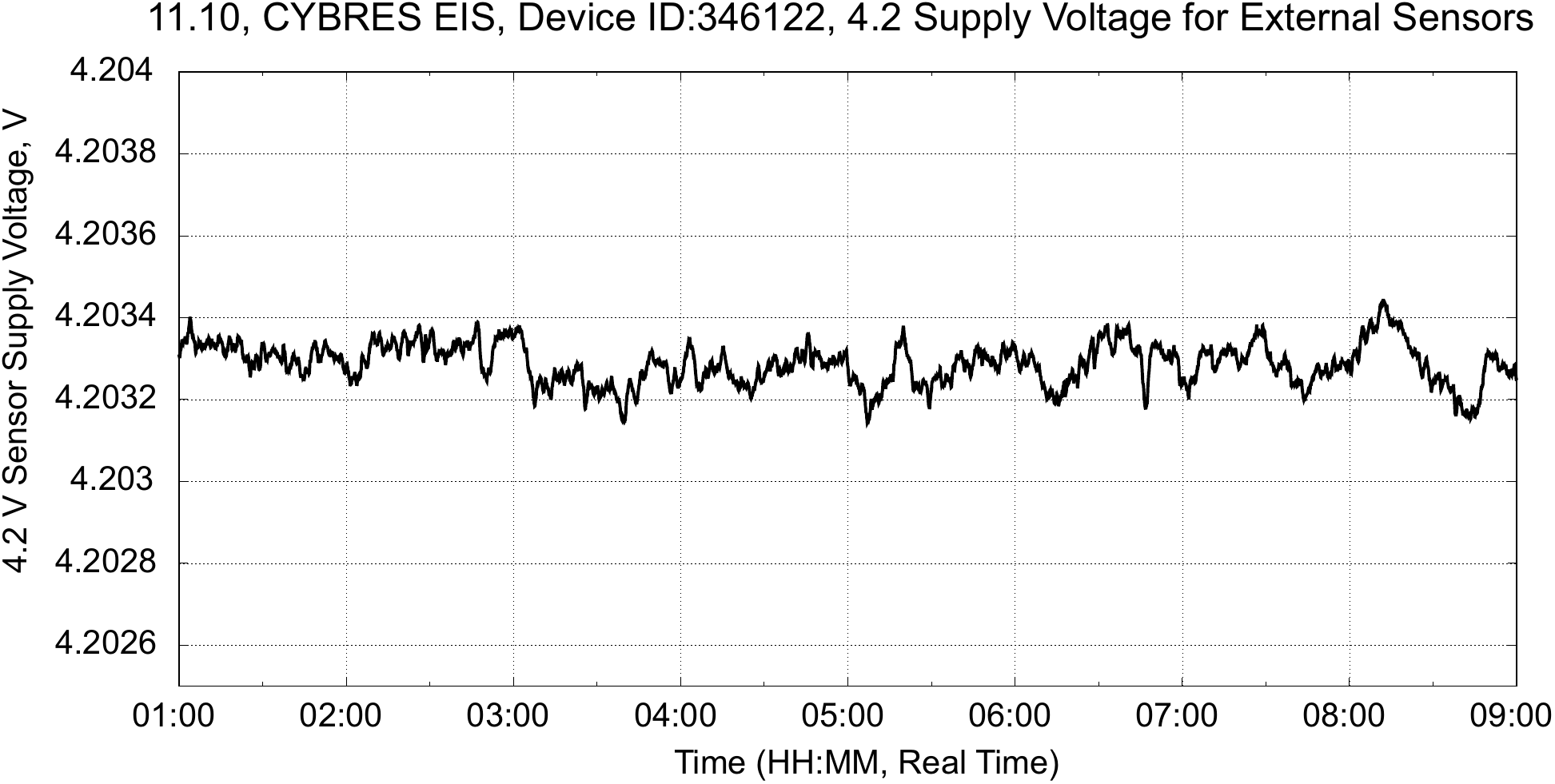}}
\caption{\small Example of experiments in the phase 2: \textbf{(a)} differential temperature of fluidic and air sensors; \textbf{(b)} temperature dynamics in measurement laboratory; \textbf{(c)} convection-based fluctuations of temperature in measurement laboratory (regression analysis) measured by external air sensors; \textbf{(d)} dynamics of power supply for fluidic and air sensors. Grey bar shows an experimental session. Other examples of fluidic and environmental data are shown in Figs.\ref{fig:phase2Env1}-\ref{fig:phase2Env3}. \label{fig:typicalTemperature}}
\end{figure}

\begin{figure*}[htp]
\centering
\subfigure[\label{fig:phase2ExpControl1}]{\includegraphics[width=0.33\textwidth]{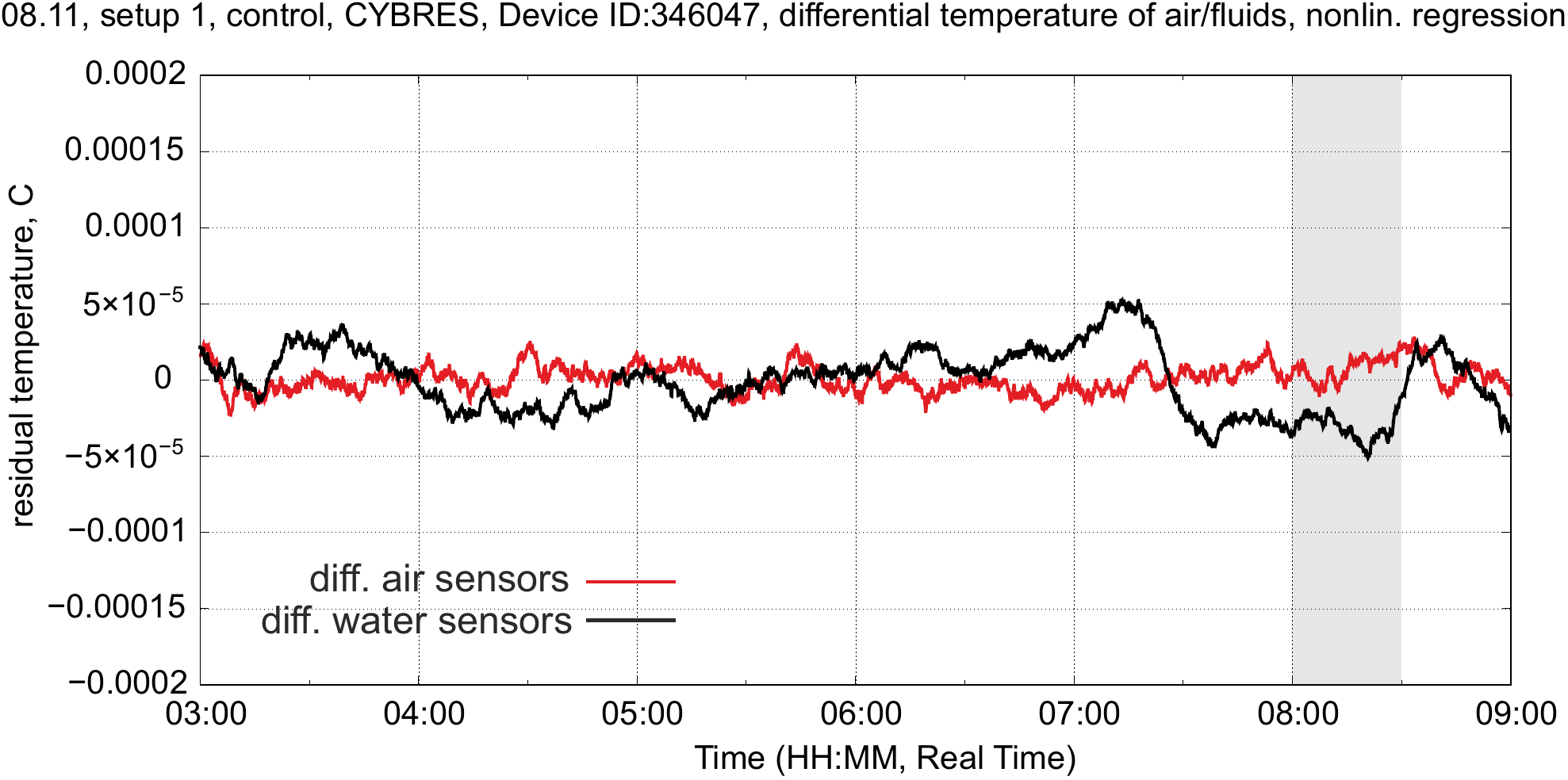}}
\subfigure[\label{fig:phase2ExpControl2}]{\includegraphics[width=0.33\textwidth]{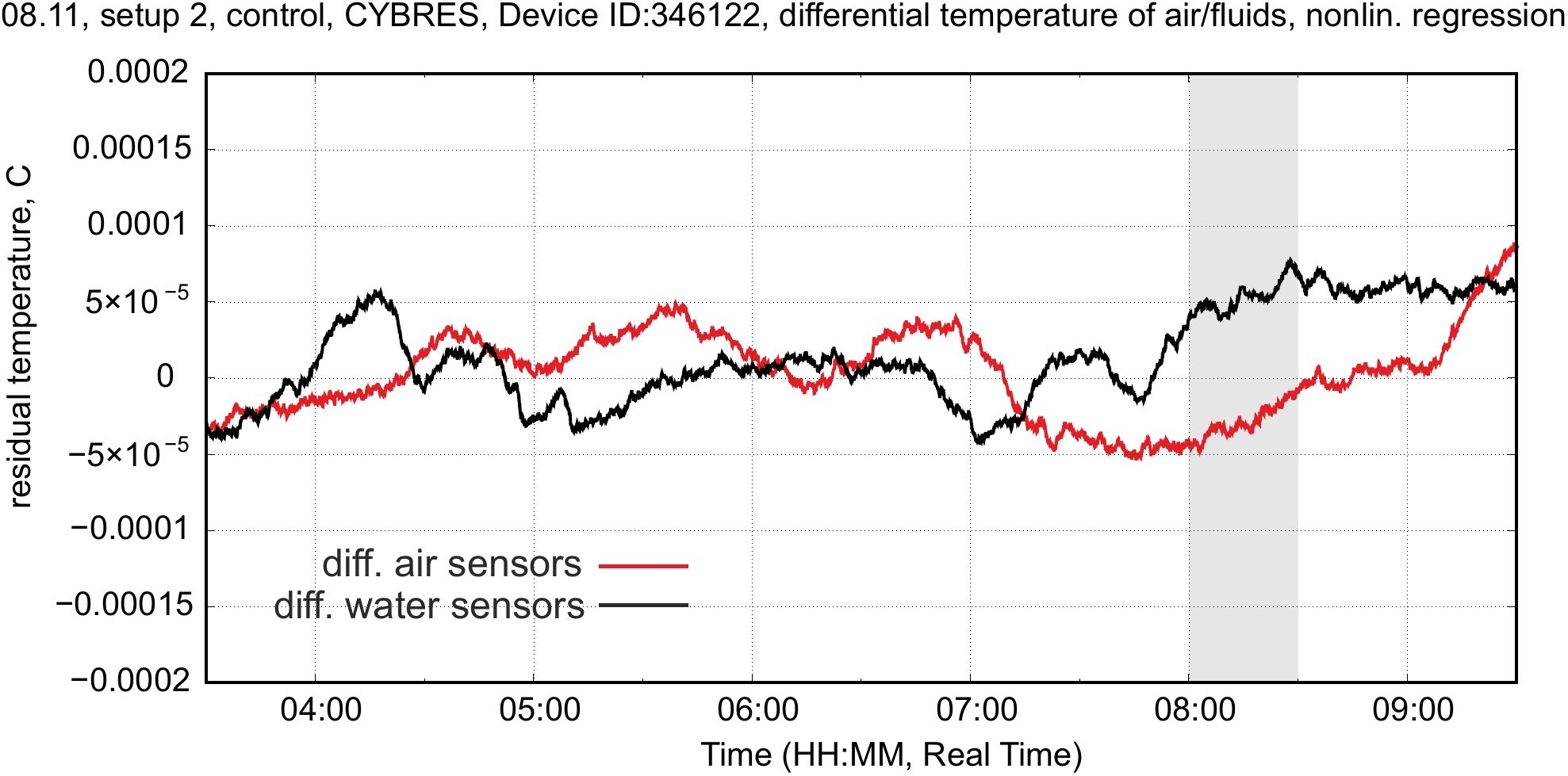}}
\subfigure[\label{fig:phase2ExpControl3}]{\includegraphics[width=0.33\textwidth]{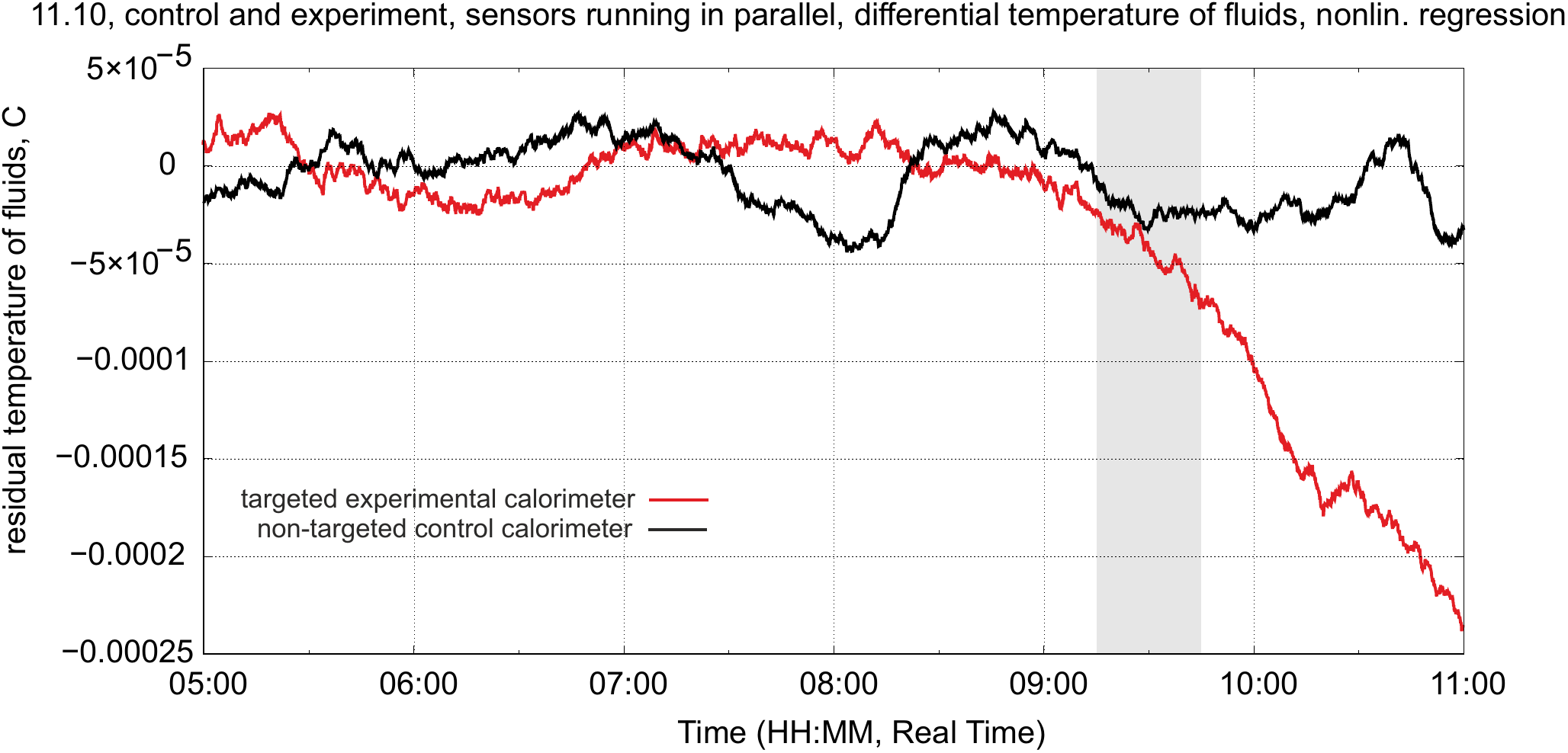}}
\subfigure[\label{fig:phase2Env1}]{\includegraphics[width=0.33\textwidth]{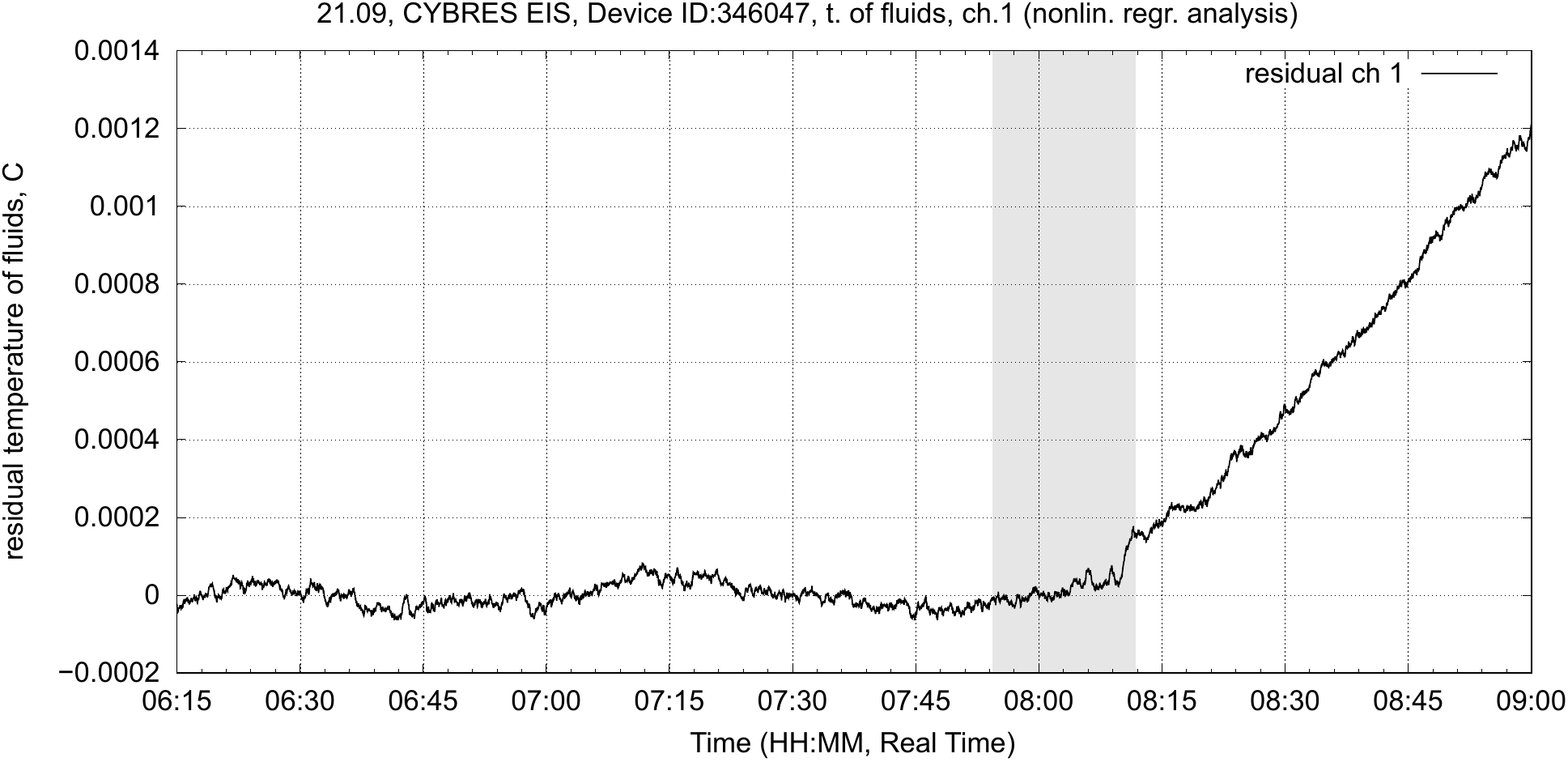}}
\subfigure[\label{fig:phase2Env2}]{\includegraphics[width=0.33\textwidth]{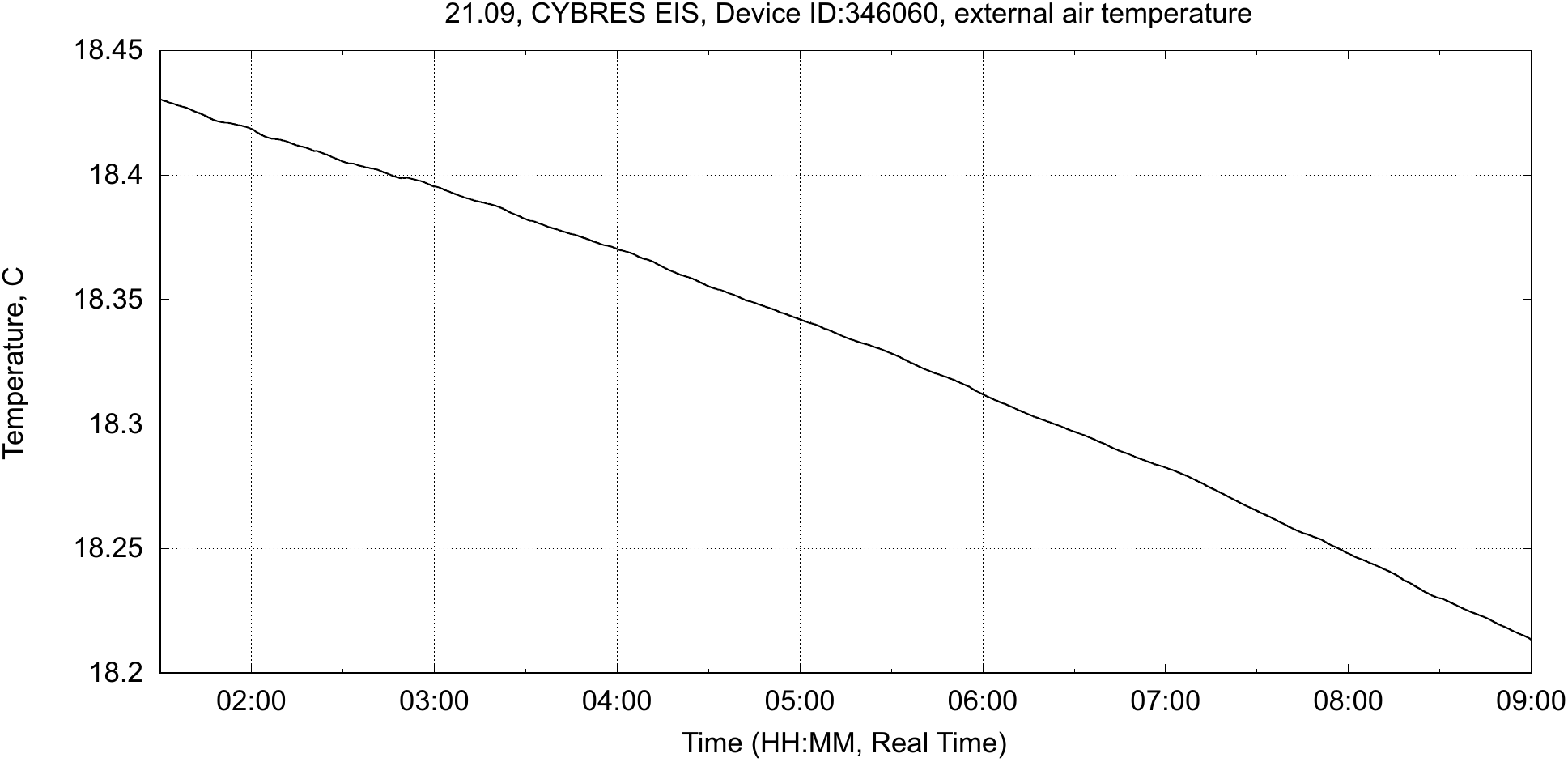}}
\subfigure[\label{fig:phase2Env3}]{\includegraphics[width=0.33\textwidth]{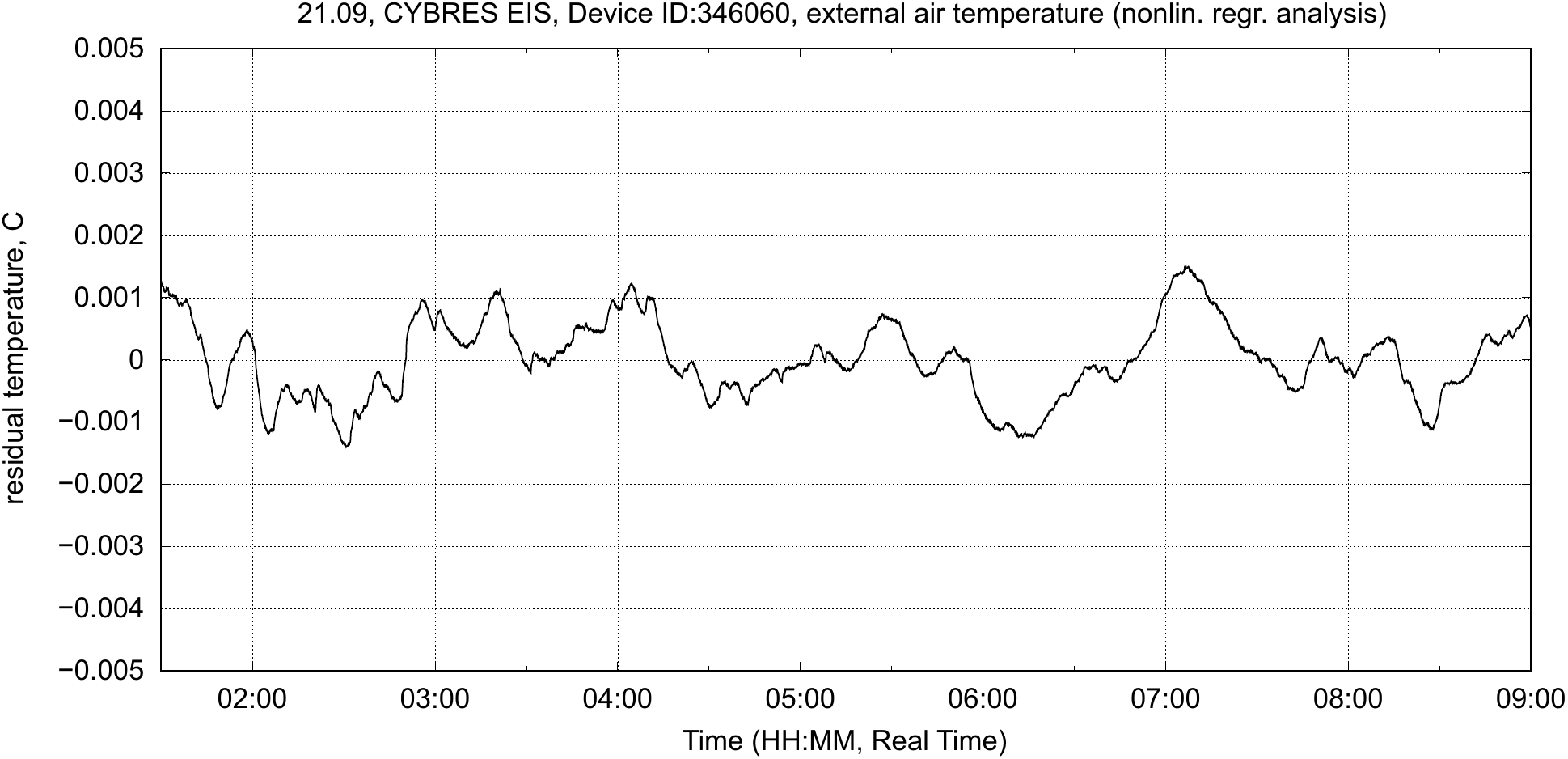}}
\subfigure[\label{fig:phase2ExpTwo}]{\includegraphics[width=0.33\textwidth]{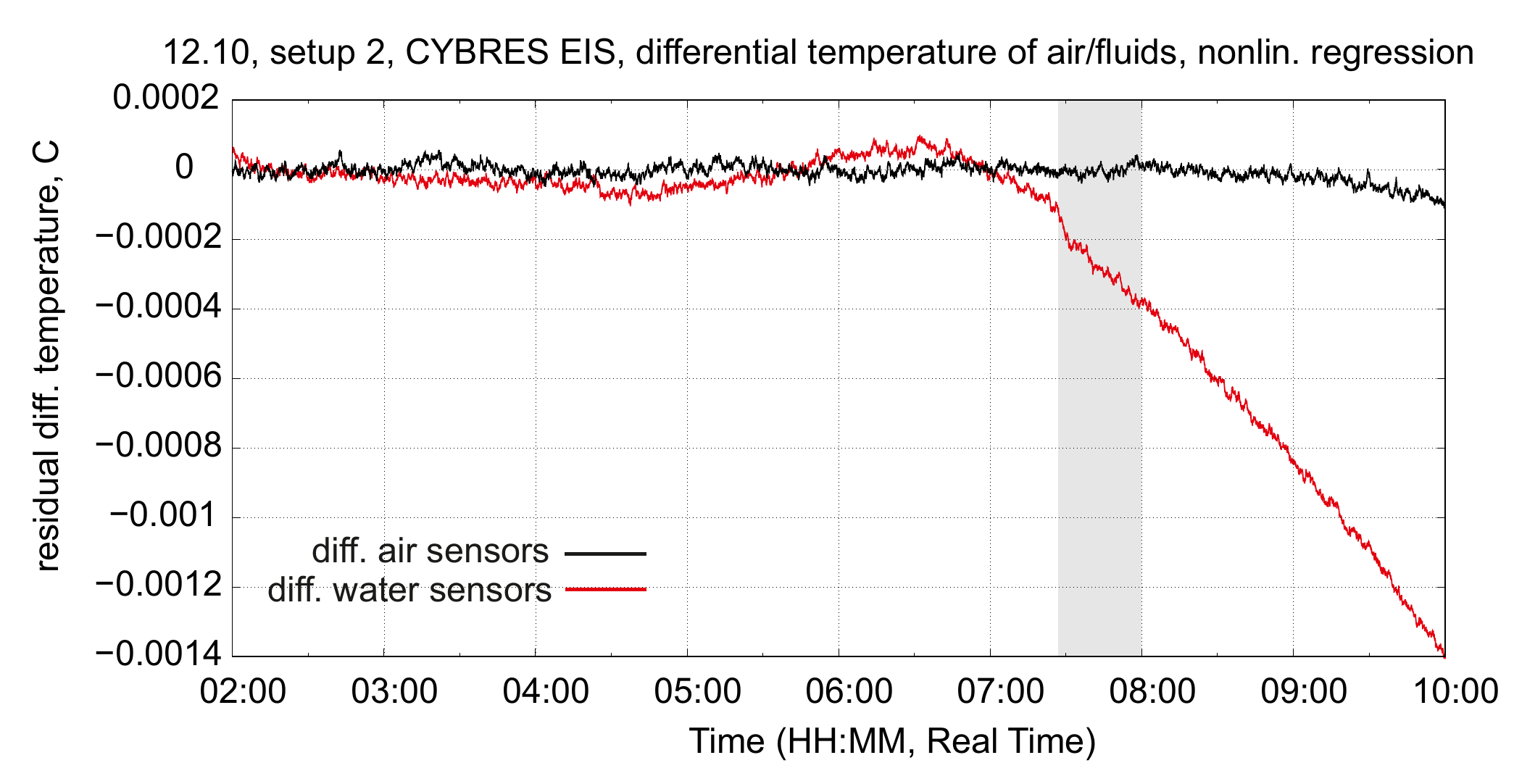}}
\subfigure[\label{fig:phase2ExpLinReg}]{\includegraphics[width=0.33\textwidth]{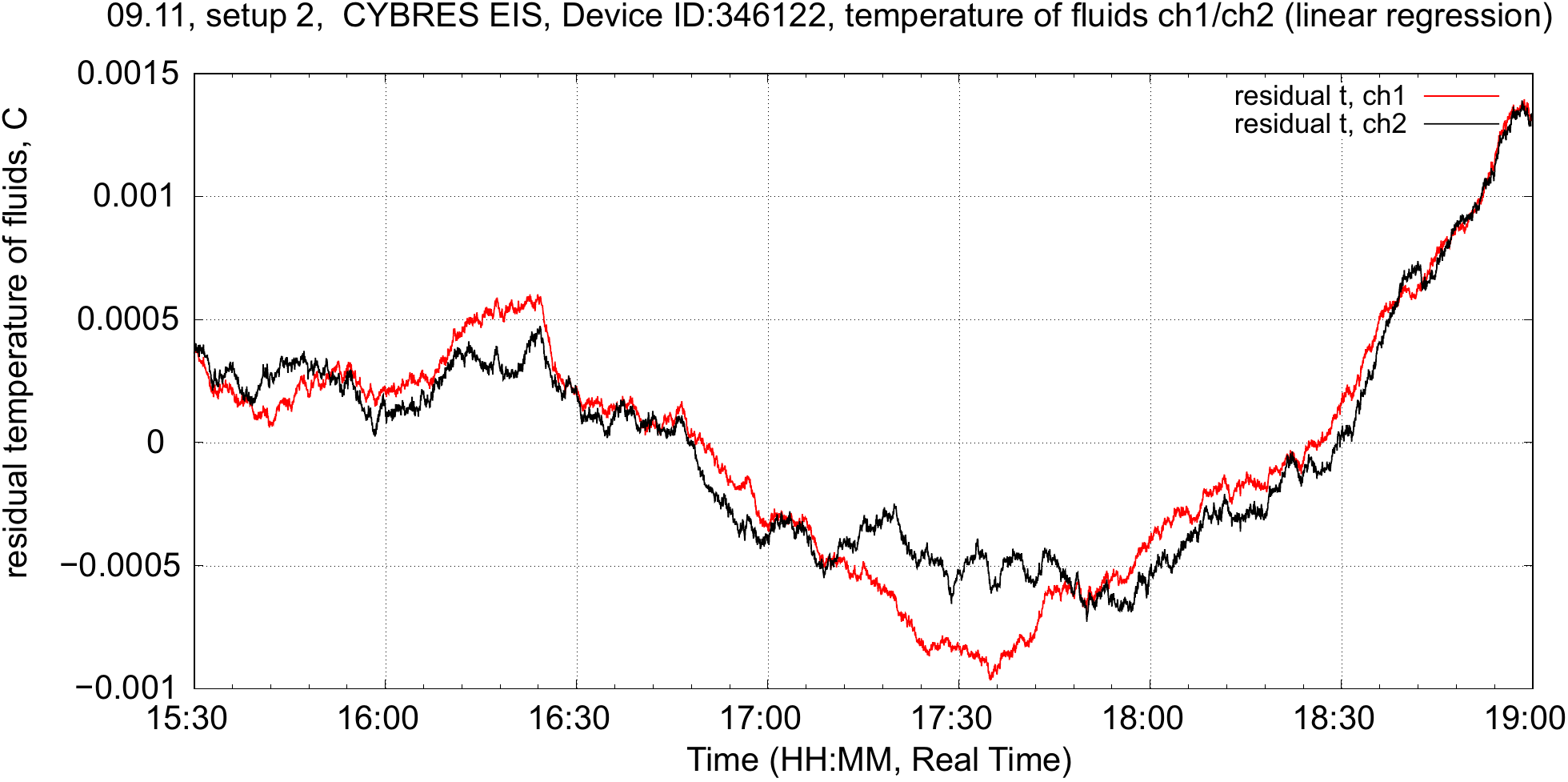}}
\subfigure[\label{fig:phase2ExpNonlinReg}]{\includegraphics[width=0.33\textwidth]{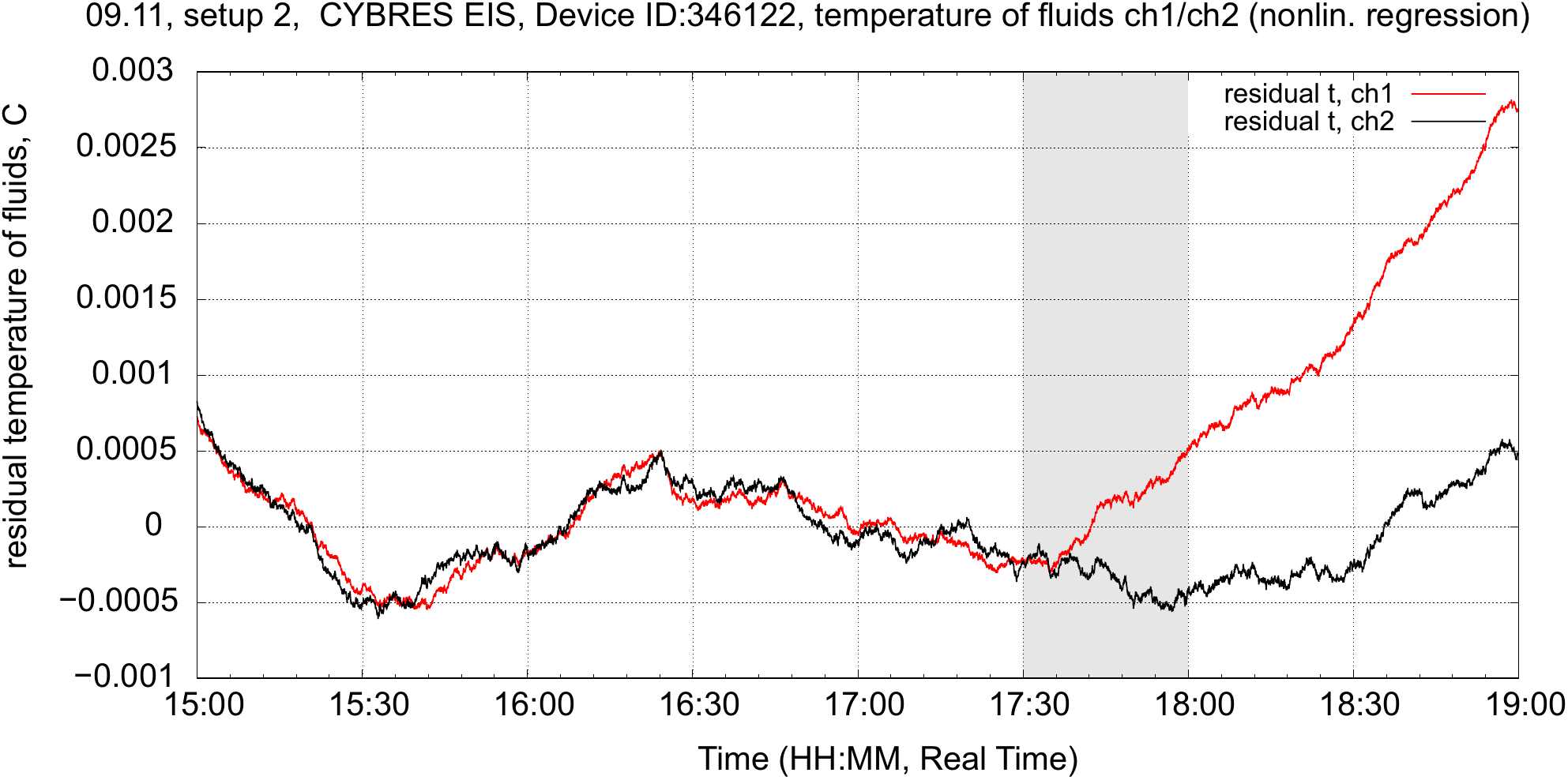}}
\subfigure[]{\includegraphics[width=0.33\textwidth]{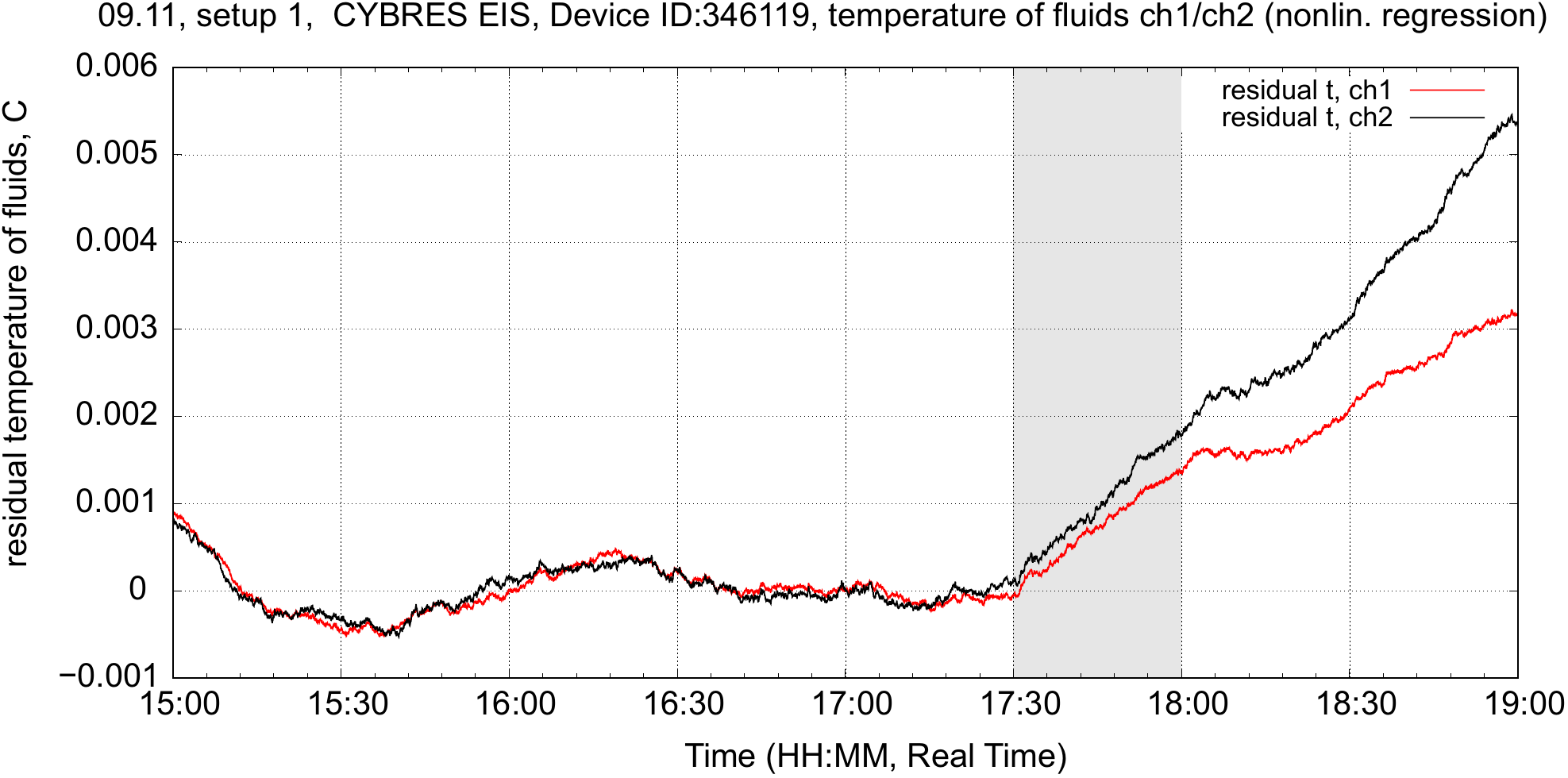}}
\subfigure[]{\includegraphics[width=0.33\textwidth]{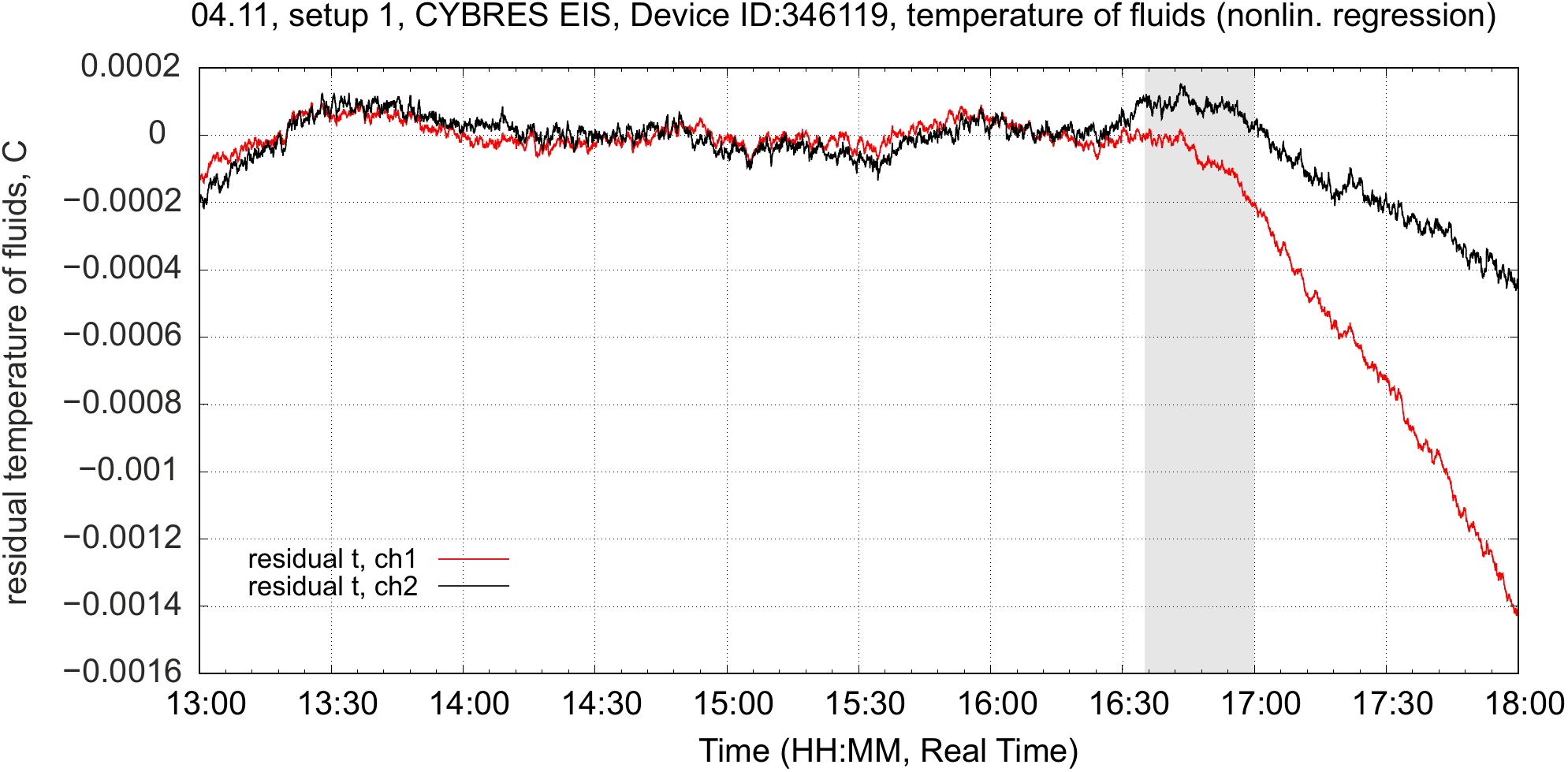}}
\subfigure[]{\includegraphics[width=0.33\textwidth]{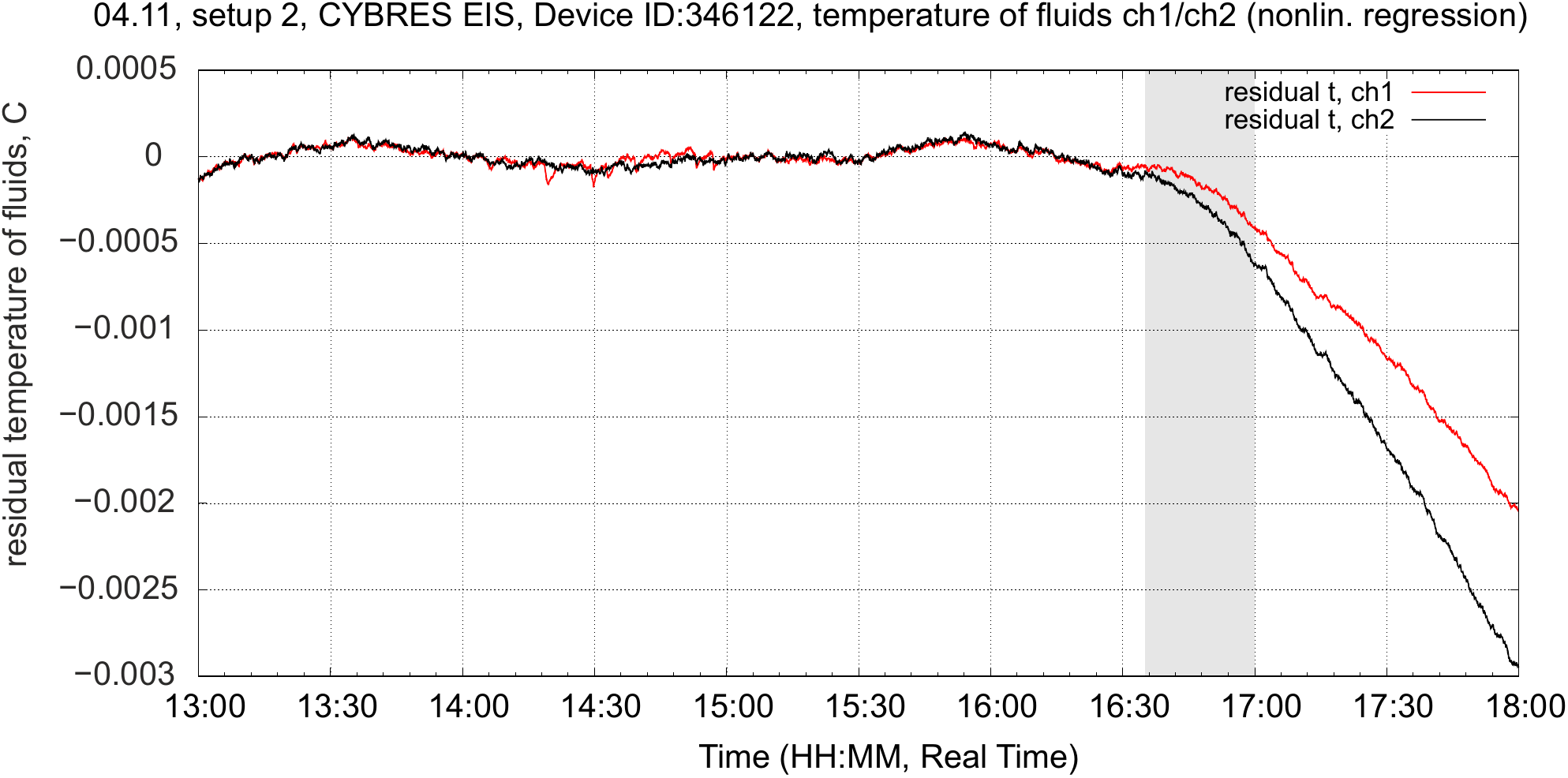}}
\subfigure[]{\includegraphics[width=0.33\textwidth]{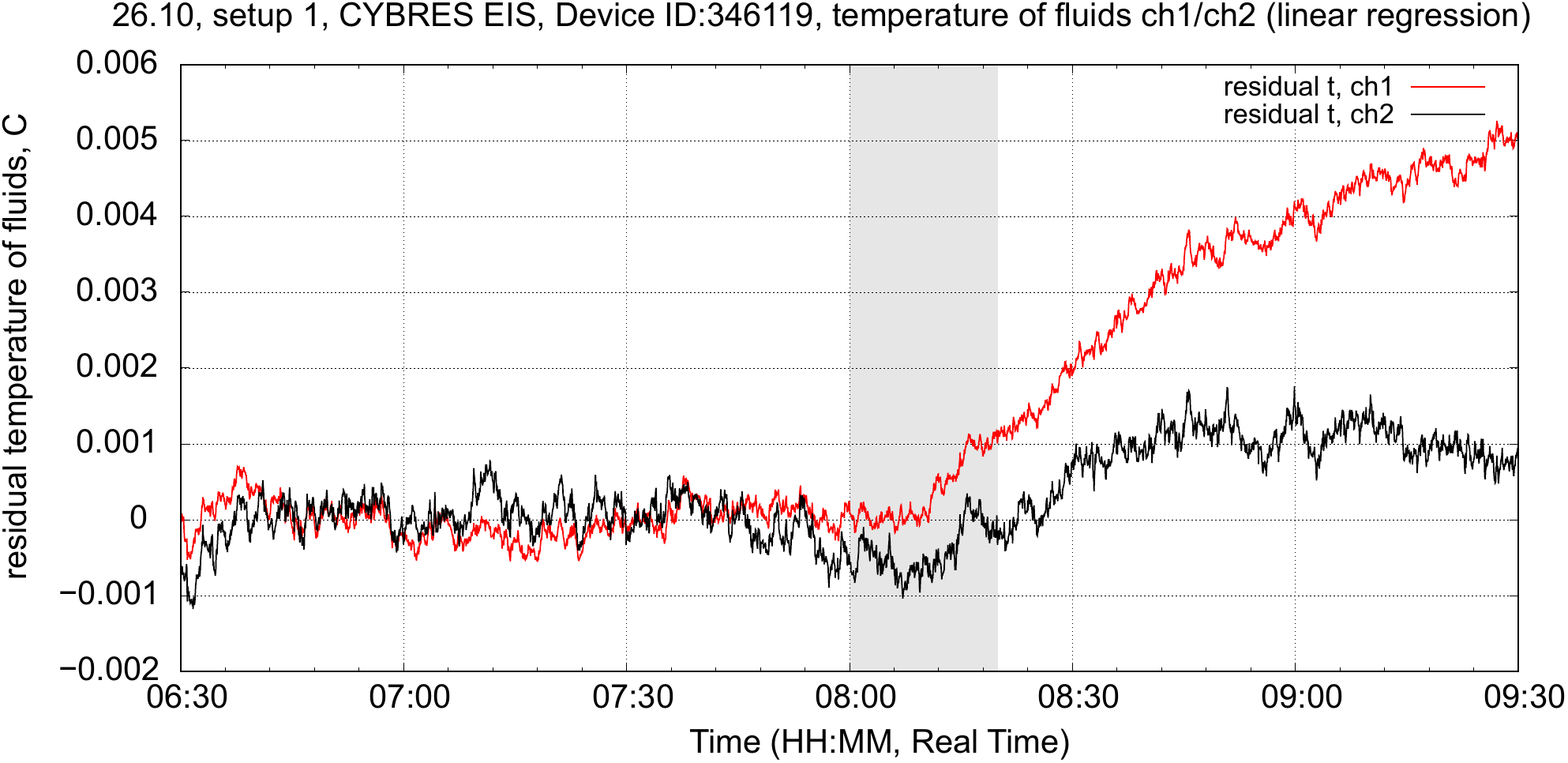}}
\subfigure[]{\includegraphics[width=0.33\textwidth]{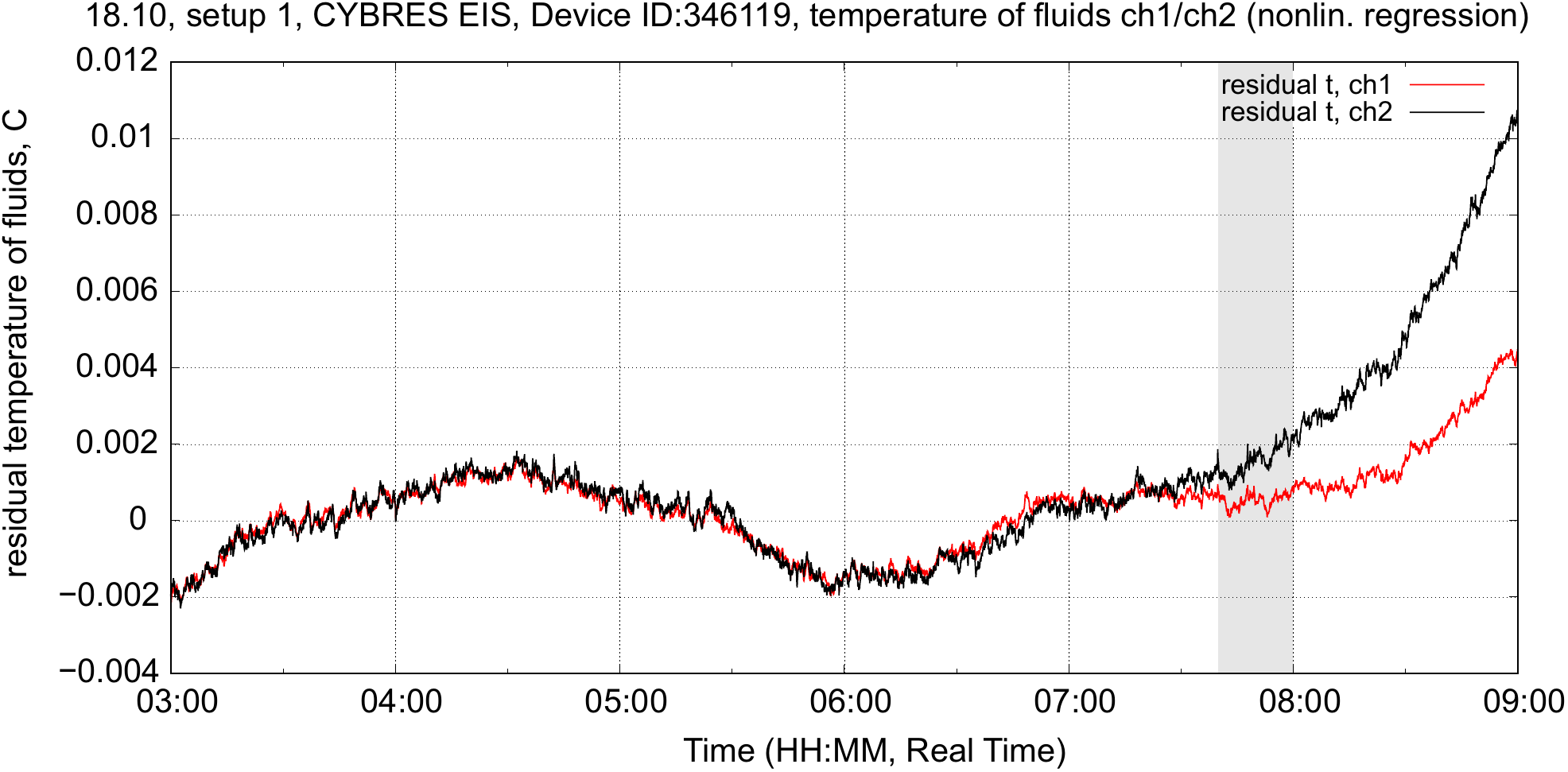}}
\subfigure[]{\includegraphics[width=0.33\textwidth]{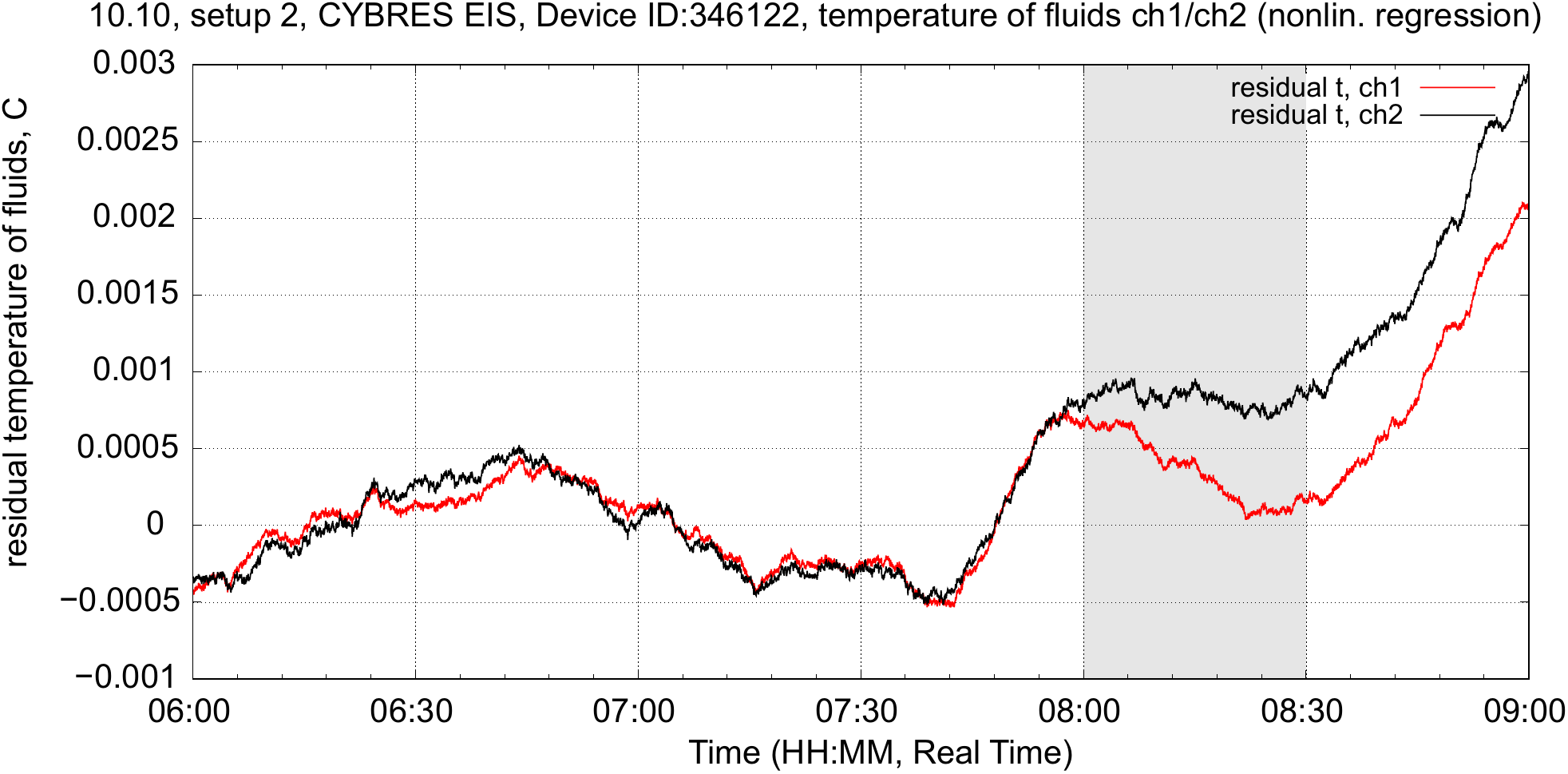}}
\subfigure[]{\includegraphics[width=0.33\textwidth]{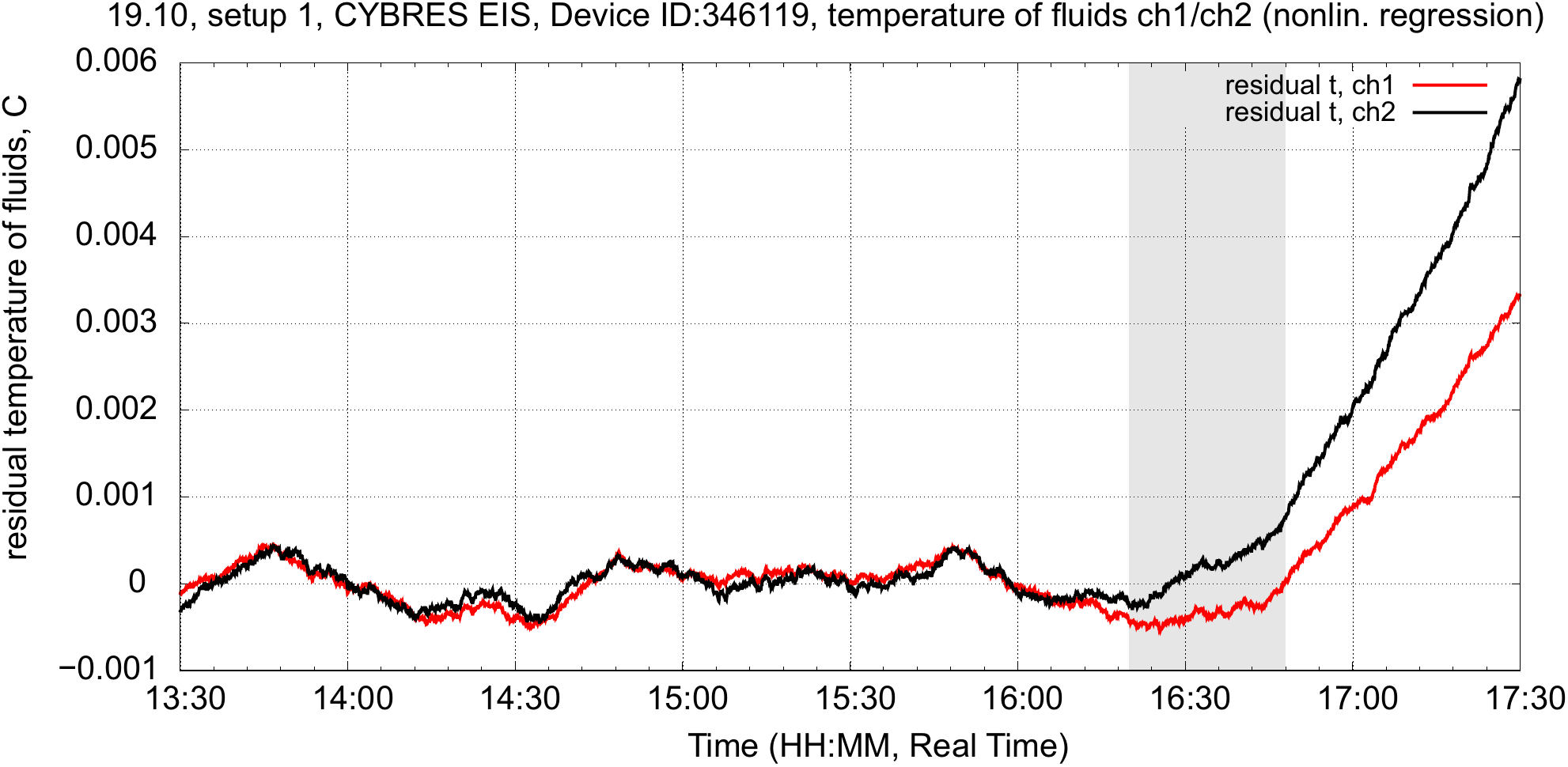}}
\subfigure[]{\includegraphics[width=0.33\textwidth]{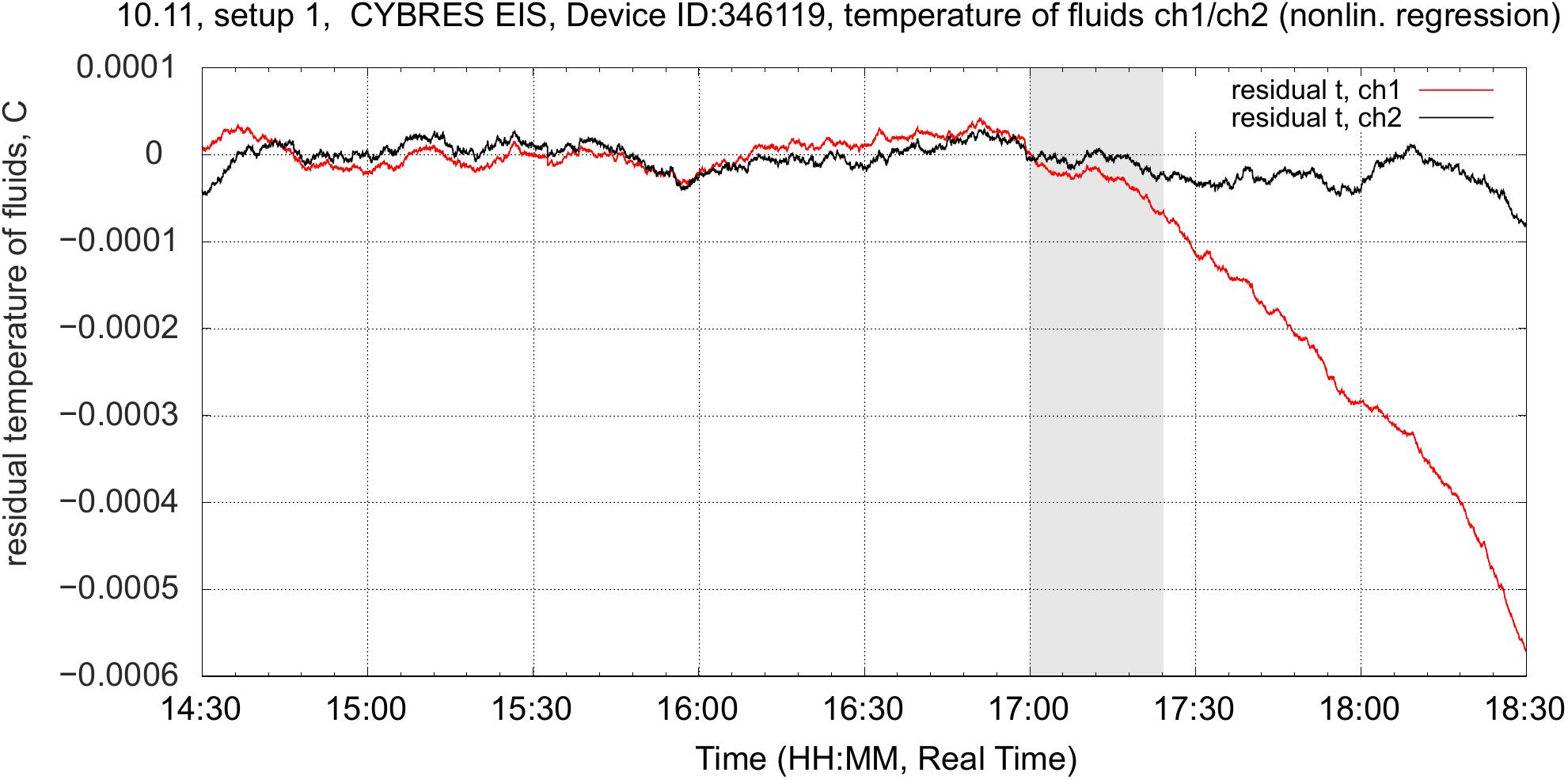}}
\subfigure[]{\includegraphics[width=0.33\textwidth]{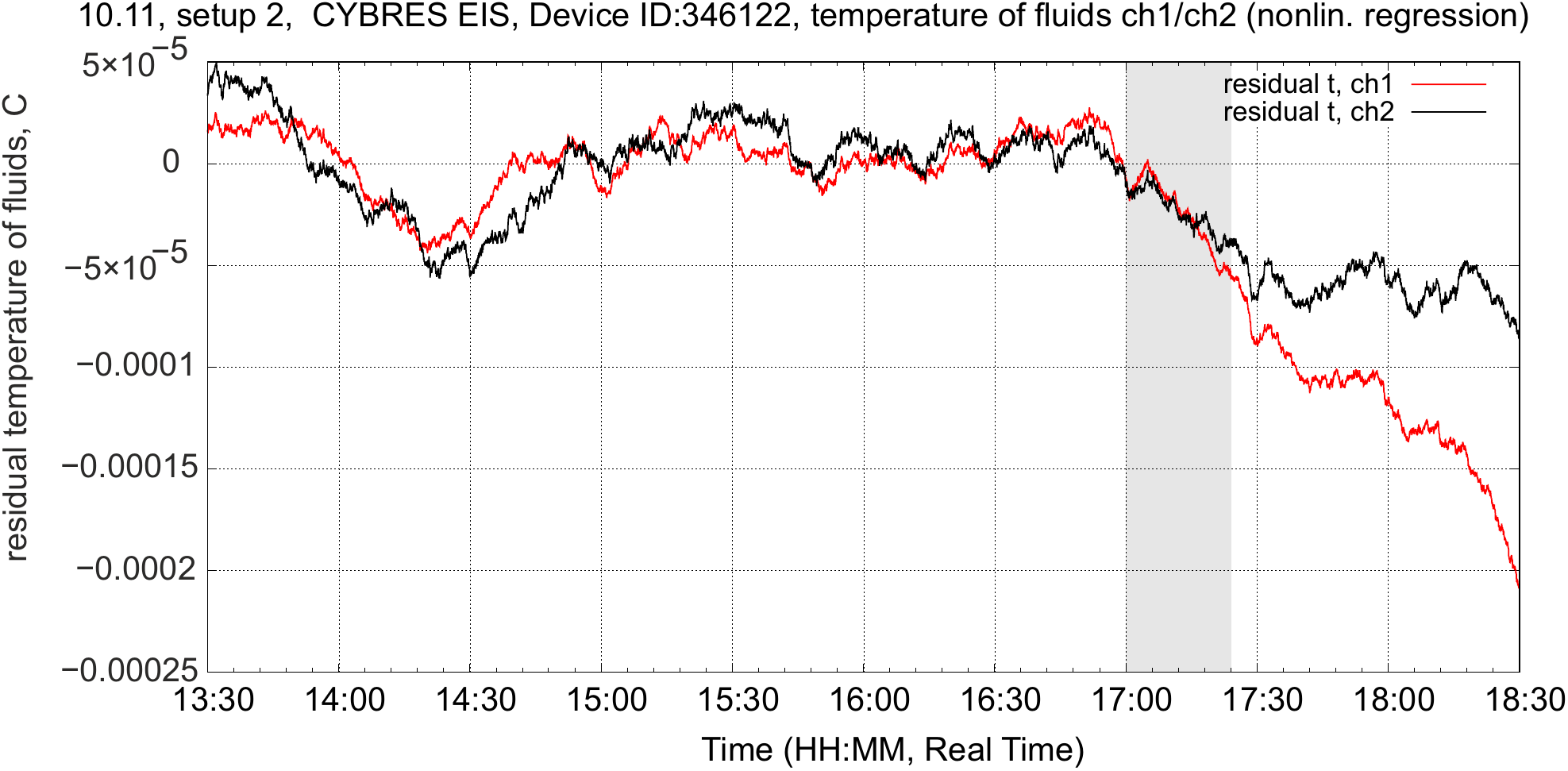}}
\caption{\small Examples of experimental data from the phase 2: \textbf{(a,b)} control measurements as 'empty sessions'; \textbf{(c)} control measurements as 'non-targeted sensors' running in parallel to experiments; \textbf{(d,e,f)} another example for environmental data as in Fig. \ref{fig:typicalTemperature}; \textbf{(a)} the same as in Fig.\ref{fig:typicalCalorimeter} but in differential form; \textbf{(h,i)} the same experiment in linear and nonlinear regression; \textbf{(j-m)} examples of multiple sensor responses; \textbf{(n-r)} other examples. \label{fig:phase2Exp}}
\end{figure*}

Fig. \ref{fig:typicalTemperature} demonstrates differential temperature of fluidic and air sensors, temperature dynamics in laboratory before, during and after the experiment (about 0.05 $^\circ$C for 8 hours, convection-based fluctuations 0.001 $^\circ$C), and dynamics of power supply for all sensors. It is clearly evident that external and system-internal factors do not provide any reasons for a change of temperature trend inside the calorimeter (especially for only one channel of the calorimeter). Experimental sessions take 20-30 minutes, temperature changes extend far beyond the sessions, typically 60-90 minutes after, until the differential temperature again balances in the calorimeters. To exclude computational artifacts caused by nonlinear regression, Figs. \ref{fig:phase2ExpLinReg} and \ref{fig:phase2ExpNonlinReg} demonstrate the same experiment in linear (without a priory knowledge about the session time) and nonlinear regression (with a priory knowledge about the session time): typically nonlinear approximation provides better fitting of background region and more clear separation of different trends, however it should not be used for long post-experimental regions due to accumulation of errors.   

Several examples of experimental data from the phase 2 are shown in Fig. \ref{fig:phase2Exp}. Since experiments involved several calorimeters working in parallel (served also as control devices), mediators attempted to affect several of them during one session. We noticed two atypical behaviors of calorimeters installed in the same measurement room. First, if at the beginning of experiments only one targeted sensor/channel responded, as shown in Fig. \ref{fig:typicalCalorimeterA}, towards the end of all experiments several fluidic sensors may have responded (however, still with a difference between targeted and not targeted channels). Second, as the experiments progressed, the water used in calorimeters exhibited increasing thermal fluctuations and therefore required more frequent water changes in the calorimeter. 

In such conditions, from 63 independent sessions conducted during three months in the phase 2, 57 attempts are positive, thus the success rate is about 90.4\%. From 67 control attempts, performed as  'empty sessions' (meditators were present but no sessions were conducted, see Figs. \ref{fig:phase2ExpControl1}, \ref{fig:phase2ExpControl2}) and in parallel to experimental sessions (as non-targeted sensors, see Fig. \ref{fig:phase2ExpControl3}), we recorded 11 positive responses mostly from non-targeted sensors running in parallel to experiments, which represent 16.4\%. The non-parametric Chi-square and Mann-Whitney U tests are used for statistical analysis against two different null-hypotheses, see also \cite{Kernbach12JSE}. For Chi-square test the positive/negative results in experimental attempts are represented by '1'/'0' and we consider the null hypothesis about a random character of obtained data (50\% of success under the experimental channel). The Mann-Whitney U test analyzes two groups of reactions from experimental and control attempts, where the null hypothesis is an identical distribution function of these groups. Results are shown in Table \ref{tab:parameters}; based on obtained values we reject the null hypotheses for Chi-square and Mann-Whitney U tests (significance level $\alpha=0.005$, two tailed).

\section{Discussions}

\textbf{1. Explanation of results.} Main discussion is related to explanation of long-term sustained temperature variations. While heating or cooling of body regions during experiments are primarily generated by somatic factors (e.g. the blood circulation intensively impacts the heat distribution in the body), it is doubtful that prolonged continued heating or cooling in post-experimental period can also be explained by these factors. Experiments with external calorimetric measurements clearly demonstrate the same long-term sustained temperature variations where somatic factors are fully excluded.

Publications with spin isomers of water suggest that temperature variations in aqueous solutions can be explained by different heat capacity of para- and ortho- isomers \cite{kernbach23Thermal}. The spin conversion changes the equilibrium 3:1 rate of spin isomers and thus changes the heat capacity. This in turn results in a symmetry breaking dynamics of temperature in differential calorimeters. The spin conversion can be triggered by different weak interactions (with minimal energy about $10^{-27}$J), e.g. by long-range spin-spin interactions \cite{Trukhanova20,Hunter:2013hza} or environmental factors \cite{KernbachPh2022}. Taking into account recent publications on non-classical brain functions, in particular on entanglement-mediated consciousness \cite{Kerskens_2022}, as a hypothesis, it is assumed that such a spin conversion induced through ASC, can cause the observed thermal effects.

We note that meditative visualization exhibits different phenomena (not only thermal effects), as shown in Fig. \ref{fig:monksData}. We see electrochemical changes (variable electrochemical reactivity caused by spin conversion \cite{kernbach2022electrochemical}) indicating begin and end of sessions, where thermal changes from body heat follow 20 minutes after electrochemical responses. It is also related to different reaction patterns in the begin (only targeted sensors) and end of experiments (several sensors at the same time). It seems that non-classical functions of the brain have different appearances, some of them are still not fully explored. For instance, positive reactions from non-targeted control experiments are described in the community \cite{Tang18}; various authors have linked them to entanglement phenomena in macroscopic systems \cite{PhysRevA.95.062116, PhysRevA.94.053807, thomas2021entanglement}.   

Involvement of spin phenomena in quantum biology \cite{10.3389/fnmol.2017.00366,Cao20} is intensively discussed in the community; they are proposed as possible candidates for consciousness-based mechanisms. Spin effects were found in biochemical reactions and cognitive processes of the brain involving phosphorus and xenon \cite{Chen20,10.1097/ALN.0000000000002226}, in various processes of animate and inanimate nature \cite{Pishchalnikov13,quantum3010006}. Natural materials (e.g., molecules) are capable of performing certain calculations at the molecular or spin levels, which is studied in natural computing \cite{Brabazon15, Lloyd06, Vedral10}. Due to long-range spin-spin interactions \cite{Trukhanova20,Hunter:2013hza}, internal and external thermal phenomena can be considered as manifestations of one common process. This new understanding allows for comprehensive exploration of ASC with modern neurocognitive and spin-based instrumentation.

\begin{figure}[htp]
\centering
\subfigure{\includegraphics[width=0.49\textwidth]{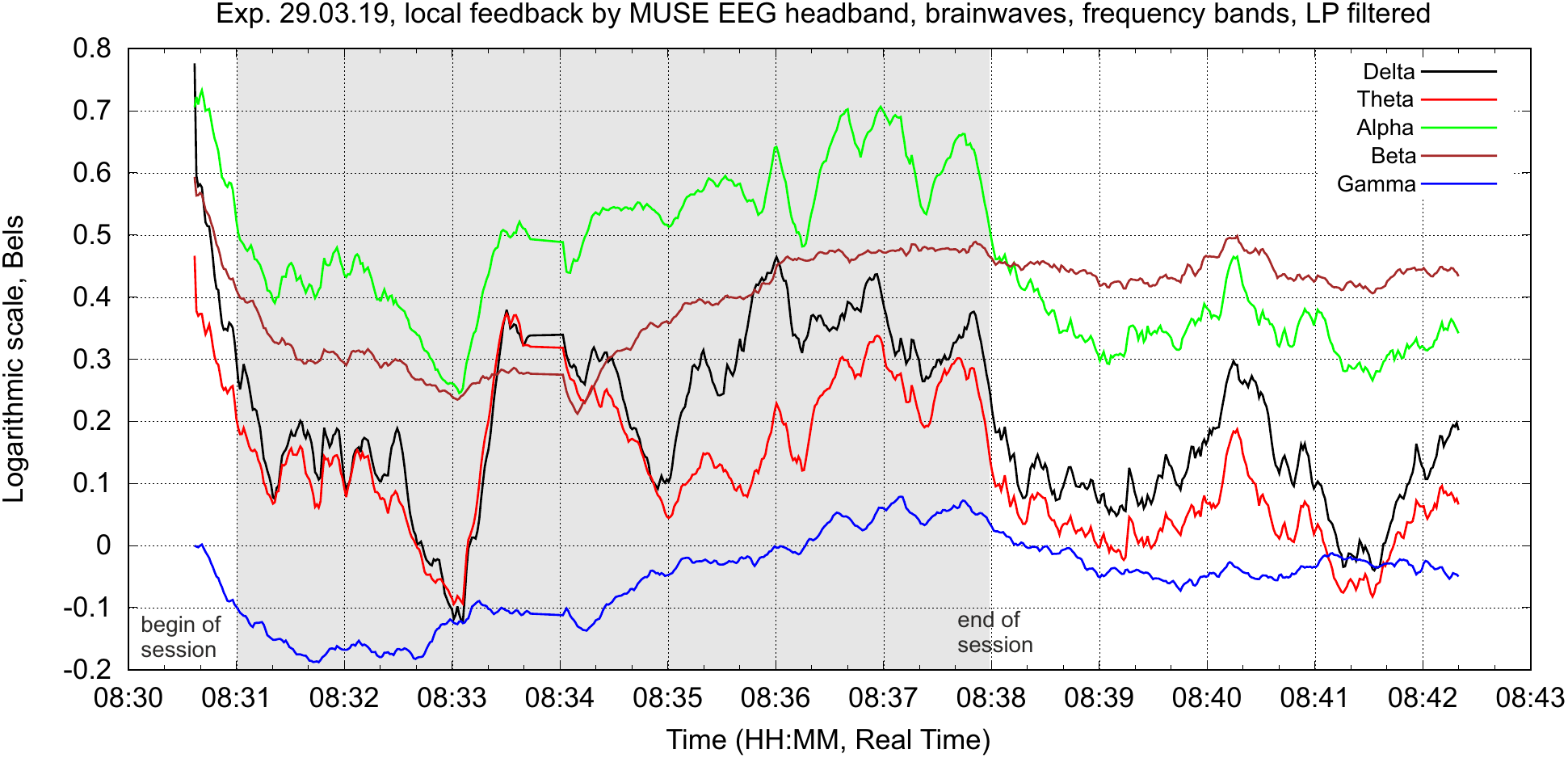}}
\subfigure{\includegraphics[width=0.49\textwidth]{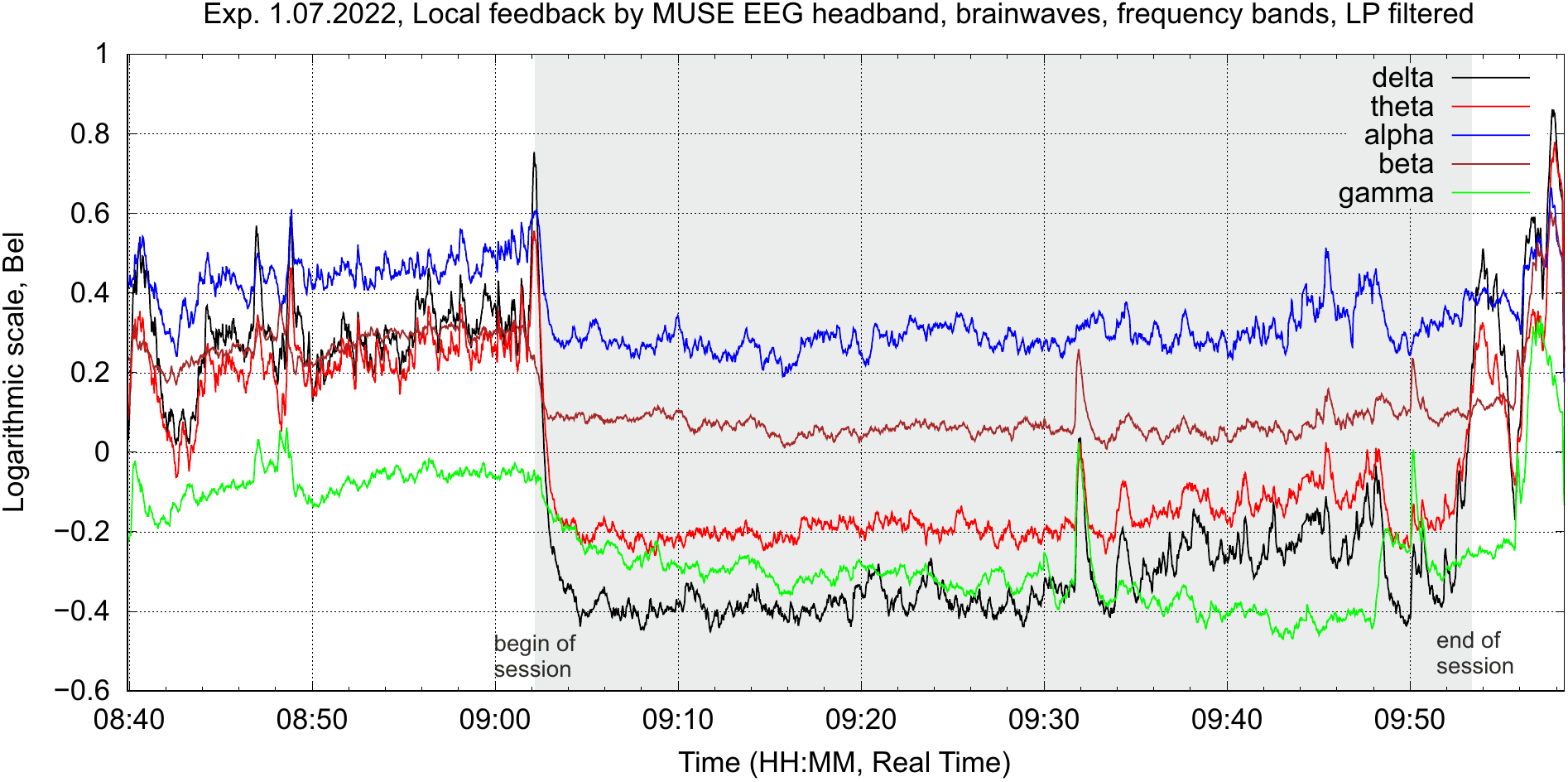}}
\caption{\small Training of meditative visualization: EEG dynamics of a meditator from 2019 and 2022 with the same exercise. \label{fig:EEG_comp}}
\end{figure}

\textbf{2. Training of operators} is already discussed in \cite{KernbachOperator19nq}. For instance, Fig.\ref{fig:EEG_comp} compares EEG of the same meditator from 2019 and 2022 during a similar active meditation exercise. It confirms the already expressed idea that periodical EEG and EIS feedback-based training positively influences the depth and duration of ASC as well as effects of related 'internal' and 'external' techniques.

\section{Conclusion}

These attempts confirm the persistent effects reported in \cite{Kozhevnikov13}. The observed variations of core and skin temperature can be interpreted as a combination of psychosomatic or biophysical mechanisms. Long-term sustained temperature effects point to biophysical processes in aqueous solutions, since they have the same appearance in water systems outside the body, such as differential calorimeters. Whereas the external thermal fluctuations can be explained by a variable heat capacity caused by spin conversion of para- and ortho- isomers of water, thermal changes in organic tissues can involve more complex spin-based processes, known in quantum biology.

The exact mechanisms of non-classical brain functions are currently under exploration in scientific community. However, on the experimental level we see a clear dependency between neurocognitive modeling in the form of meditative visualization and thermal effects at least in three different types of experimental setups. The largest series of experiments were carried out with high-resolution differential calorimeters and replicate results of three other laboratories \cite{Gurtovoi92en, Xin99, Godik10} with similar outcomes. From our point of view, the most important factor in these attempts represents a systematic training -- corresponding changes are well visible on EEG as more deeper ASC of trained meditators and their better results.

To some extent focusing attention on neurohumoral regulative centers of body can be viewed as supporting (or affecting) their functionality by consciousness-based mechanisms in ASC. Such an interpretation requires further biological and medical research, but it can explain various cases of healing and demonstration of unusual abilities of the body (for example, pouring water on the body in severe frost in the Himalayas in Tummo yoga \cite{Kozhevnikov13}).

\section{Acknowledgment}

We thank Phra Pisit Suksuthipunth and Phra Puttipong Eakwong for multiple discussions about the Dhammakaya meditation. Special thanks to the Buddhist Meditation Center in Heilbronn for providing meeting and meditation opportunities.

\small
\IEEEtriggeratref{43}


\end{document}